\documentclass[useAMS,usenatbib]{mn2e} 
\usepackage{graphicx}
\usepackage{color}
\usepackage{txfonts}
\usepackage{epstopdf}

\title[Globular clusters with Gaia]{Globular clusters with
Gaia}

\author[Pancino et al.]{E. Pancino$^{1,2}$\thanks{email:pancino@arcetri.inaf.it}, 
                        M. Bellazzini$^{3}$, 
                        G. Giuffrida$^{4,2}$, 
                        S. Marinoni$^{4,2}$\\  
$^{1}$INAF-Osservatorio Astrofisico di Arcetri, Largo Enrico Fermi 5, 50125,
Firenze, Italy\\
$^{2}$ASI Science Data Center, Via del Politecnico SNC, I-00133 Rome, Italy\\
$^{3}$INAF-Osservatorio Astronomico di Bologna, Via Ranzani 1, 40127, Bologna, 
Italy\\
$^{4}$INAF-Osservatorio Astronomico di Roma, Via di Frascati 33, 00044
Monteporzio Catone, Italy}

\begin{document}

%

\date{\today}

\pagerange{\pageref{firstpage}--\pageref{lastpage}} \pubyear{2013}

\maketitle
\label{firstpage}

\begin{abstract} 

The treatment of crowded fields in Gaia data will only be a reality in a few
years from now. In particular, for globular clusters, only the end-of-mission
data (public in 2022--2023) will have the necessary full crowding treatment and
will reach sufficient quality for the faintest stars. As a consequence, the work
on the deblending and decontamination pipelines is still ongoing. We describe
the present status of the pipelines for different Gaia instruments, and we model
the end-of-mission crowding errors on the basis of available information. We
then apply the nominal post-launch Gaia performances, appropriately worsened by
the estimated crowding errors, to a set of 18 simulated globular clusters with
different concentration, distance, and field contamination. We conclude that
there will be 10$^3$--10$^4$ stars with astrometric performances virtually
untouched by crowding (contaminated by $<$1~mmag) in the majoritiy of clusters.
The most limiting factor will be field crowding, not cluster crowding: the most
contaminated clusters will only contain 10--100 clean stars. We also conclude
that: (i) the systemic proper motions and parallaxes will be determined to 1\%
or better up to $\simeq$15~kpc, and the nearby clusters will have radial
velocities to a few km~s$^{-1}$ ; (ii) internal kinematics will be of
unprecendented quality, cluster masses will be determined to $\simeq$10\% up to
15~kpc and beyond, and it will be possible to identify differences of a few
km~s$^{-1}$ or less in the kinematics (if any) of cluster sub-populations up to
10~kpc and beyond; (iii) the brightest stars (V$\simeq$17~mag) will have
space-quality, wide-field photometry (mmag errors), and all Gaia photometry
will have 1--3\% errors on the absolute photometric calibration.

\end{abstract}


\section{Introduction}

The ESA\footnote{A list of the acronyms used in this paper
(Table~\ref{tab:acr}) can be found in Appendix~\ref{sec:acr}.} (European Space
Agency) space mission Gaia\footnote{http://www.cosmos.esa.int/web/gaia}
\citep{perryman01,mignard05,gaia1,gaia2} is the successor of Hipparcos
\citep{perryman97}, with the goal of providing astrometry for billions of
point-like sources across the whole sky, with an error of 24~$\mu$as-level for
stars of G$\simeq$15~mag. Gaia will also provide broadband magnitudes and
colours for all sources, down to the Gaia white-light magnitude of G=20.7~mag
(V$\simeq$21~mag). Low-dispersion spectra will also be obtained in two broad
bands with the red and blue spectrophotometers (BP and RP). Finally, Gaia will
produce spectra in the calcium triplet region with the RVS (the Radial Velocity
spectrometer) down to G$\simeq$17~mag, from which radial velocities (RVs).
Object classification and parametrization will be possible for all sources. Gaia
was launched in December 2013 \citep{debruijne14}, and the first data release
was in September 2016 \citep{gaia2}, containing positions and white-light
magnitudes for the best behaved stars, and additional information like
parallaxes and proper motions for $\simeq$2 million stars observed previously
with Tycho-2 \citep{hog00,michalik15,tgas}.

Gaia will observe not only stars, but also tens of thousands of quasars,
unresolved galaxies, Solar system objects, many transient and variable objects
like supernovae, and finally the interstellar medium
\citep{altavilla12,ducourant14,eyer14,debruijne15,proft15,zwitter15,bachchan16,tanga16}.
Gaia will also pose a challenge because of its data amount and complexity,
pushing the astrophysical community farther into the path of big data and data
mining \citep{gaia1}. 

Gaia is limited in dense stellar fields, owing to the on-board and downstream
telemetry bandwidth. For spectroscopy, an additional limitation is provided by
the large physical size of the dispersed images and spectra on the focal plane.
This has particular relevance for studies of the Galactic plane, the bulge, and
globular clusters (hereafter, GCs). In the first few Gaia data releases,
disturbed sources like binaries, multiple stars, and stars in crowded fields
will likely not be part of the released material \citep{gaia2}. In any case, the
inclusion of a sufficient sample of stars at the main sequence turn-off point or
fainter -- with good quality measurements -- is very important for GC research.
Therefore, only the latest few Gaia releases (2020--2023) are expected to
provide a significant breakthrough in GC research. To prepare the work, we
explore in this paper the expected behaviour of Gaia data in several simulated
Galactic GCs, adopting the official post-launch Gaia science performances and
some simplified recipes to describe additional deblending error components,
based on the Gaia deblending pipelines. A very preliminary -- and now outdated
-- version of this work was presented by \citet{pancino13}. 

The paper is organized as follows: in Section~\ref{sec:gaia} we describe the
current status of crowding treatment in Gaia; in Section~\ref{sec:crowderr} we
present our crowding errors modeling; in Section~\ref{sec:ggc_simu} we describe
our simulated clusters and the computation of Gaia observed quantities and final
errors; in Section~\ref{sec:res} we explore the simulations and show the
potential of Gaia data for GC studies; in Section~\ref{sec:concl} we summarize
the main results and draw our conclusions.

\begin{figure}
 \centering
 \includegraphics[width=\columnwidth,height=5.4cm]{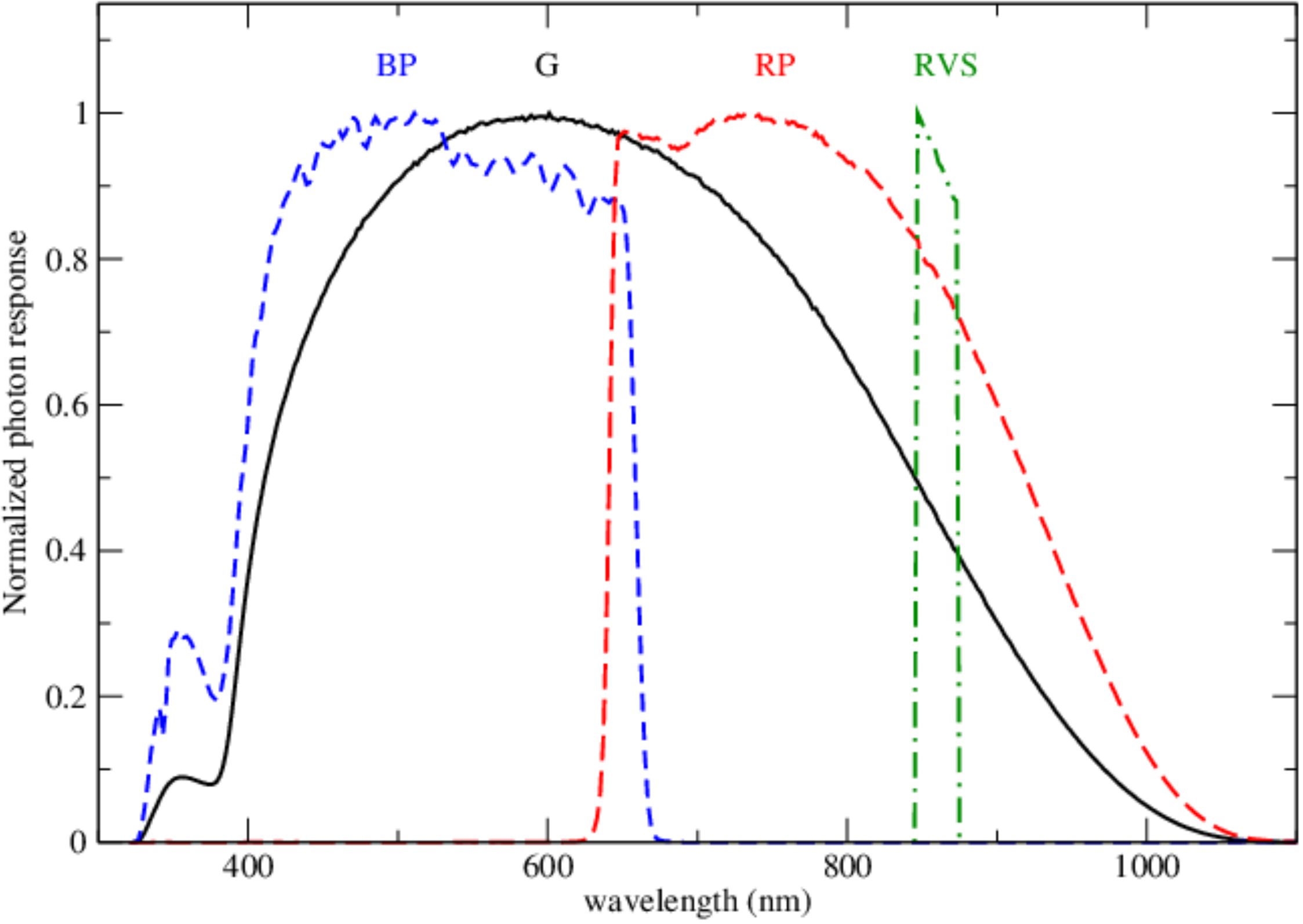}
 \caption{The Gaia nominal passbands for the astrometric field (G-band), the
 blue and red spectrophotometers (BP and RP), and the radial velocity
 spectrometer (RVS). Figure courtesy of C.~Jordi.}
  \label{fig:filters}
\end{figure}

\section{Crowding in Gaia data} 
\label{sec:gaia}

As mentioned in the previous section, Gaia is a complex space mission, with two
different telescopes projecting their light on a large common focal plane,
captured by 106 different CCDs (Charged-Coupled Devices), and passing through
different instruments. Gaia scans the whole sky by precessing its spin axis, and
describing great circles on the sky that slowly drift with its precession. Each
region is scanned from a minimum of $\simeq$40 times, to a maximum of
$\simeq$250 times, with an average number of $\simeq$70 passages for the AF, BP
and RP, and $\simeq$40 passages for RVS. All the CCDs in the focal plane are
read in TDI (Time-Delayed Integration) mode to closely follow Gaia's movement
across the sky. On the focal plane, stars ``move" along the scanning direction,
encountering different instruments:

\begin{itemize}
\item{the first two columns of the CCD array are called Sky Mappers (SM), and
they are used for the on-board detection of point-like sources; each of the SM
columns sees only light from one of the two telescopes;}
\item{the AF (Astrometric Field), provides astrometry and photometry of
point-like sources in the G-band, a white-light passband defined by the telescope
and instrument transmission and by the CCD quantum efficiency
(Figure~\ref{fig:filters});} 
\item{the BP and RP provide low resolution spectra
(R=$\lambda/\delta\lambda$=20-100) in the ranges shown in
Figure~\ref{fig:filters}, and the integrated G$_{\rm{BP}}$ and G$_{\rm{RP}}$
magnitudes; the spectra are necessary for the chromaticity displacement
correction in astrometric measurements;}
\item{the RVS, provides R$\simeq$11\,700 spectra in the calcium triplet region,
for stars down to G$\simeq$17~mag, depending on the object.}
\end{itemize}

To save telemetry bandwidth, given the enormous amount of data produced daily by
Gaia, observations in each instruments are only transmitted for pixels contained
in rectangular windows, that follow the detected point-like objects along the
focal plane. For the faint stars, data in the allocated windows are binned in
the AC direction by the on-board processing software. The adopted window sizes
and relevant quantities for our treatment of crowding are listed in
Table~\ref{tab:numbers}. More information can be found in the Gaia mission paper
\citep{gaia1}.  

\subsection{Gaia deblending and decontamination pipelines}

Crowding treatment ideally requires a preliminary evaluation of the crowding
conditions of each source and transit, based on knowledge of the {\em scene},
i.e., the distribution and characteristics of all the neighbouring sources, as
collected before the current observation. Different pipelines are employed for
different instruments and to treat different cases. Here below we describe the
current status of the ones that are relevant for the present study.

\begin{table}
\caption{Adopted Gaia relevant quantities. We note that the AF window sizes are
relevant for the astrometry; the BP/RP sizes for the photometry, chromaticity
correction, and object classification and parametrization; the RVS window sizes
are relevant for radial velocity, object parametrization and abundance analysis.
\label{tab:numbers}}
\begin{center}
\begin{tabular}{rrl}
\hline
\\
Size     & Size      & Description \\
(")      &(pix)      & \\
\hline
 0.176789 &    1 (AC) & Across scan (AC) pixel size \\
 0.058933 &    1 (AL) & Along scan (AL) pixel size \\
 0.176789 &    3 (AL) & Gaia PSF \\
 2.121468 &   12 (AC) & AC window size (AF, BP, and RP)\\
 0.070720 &   12 (AL) & AL window size (AF)$^a$ \\
 1.767890 &   10 (AC) & AC window size (RVS) \\
 3.535980 &   60 (AL) & AL window size (BP and RP) \\
74.785977 & 1269 (AL) & AL window size (RVS) \\
\hline
\end{tabular}\\
\raggedright{$^a$ In some CCD columns in AF, as well as in the SM CCD columns,
the windows are longer to allow for background measurements around the sources.
The longer wings of these windows can in principle be used also to check the
source profile behaviour ourtside normal window limits.}\\
\end{center}
\end{table}

\subsubsection{AF deblending}
\label{sec:nss}

Stars closer than the Gaia PSF width (which is assumed here\footnote{The value
we adopted is a conservative estimate. The Gaia {\em effective} PSF varies
across the field of view and has a median value of 0.103" \citep{fabricius16}.}
to be 0.177", see also Table~\ref{tab:numbers}) can be recognized as blends
already in the astrometric processing of AF data. These blends can be detected,
for example, from the high errors in the centroid determination and its wobbling
from one transit to another, or by photometric variability, radial velocity
variability, or in general because the fits of the PSF or LSF to the data show
large residuals or require more than one component. 

Assessing whether a star is isolated or has detected or suspected companions is
a crucial task. Multiple or blended objects are redirected to the NSS
(Non-Single Stars) pipeline where an attempt to model them as binary systems is
carried out \citep{gaia1,pourbaix11}. If none of the available binary models or
configurations produces a good fit to the data, then a stochastic model is
employed to derive preliminary parameters of the secondary (or tertiary and so
on) source. Therefore we can assume that -- considering also the small Gaia PSF
-- the vast majority of NSS will be known. 

\begin{figure}
 \centering
 \includegraphics[width=\columnwidth]{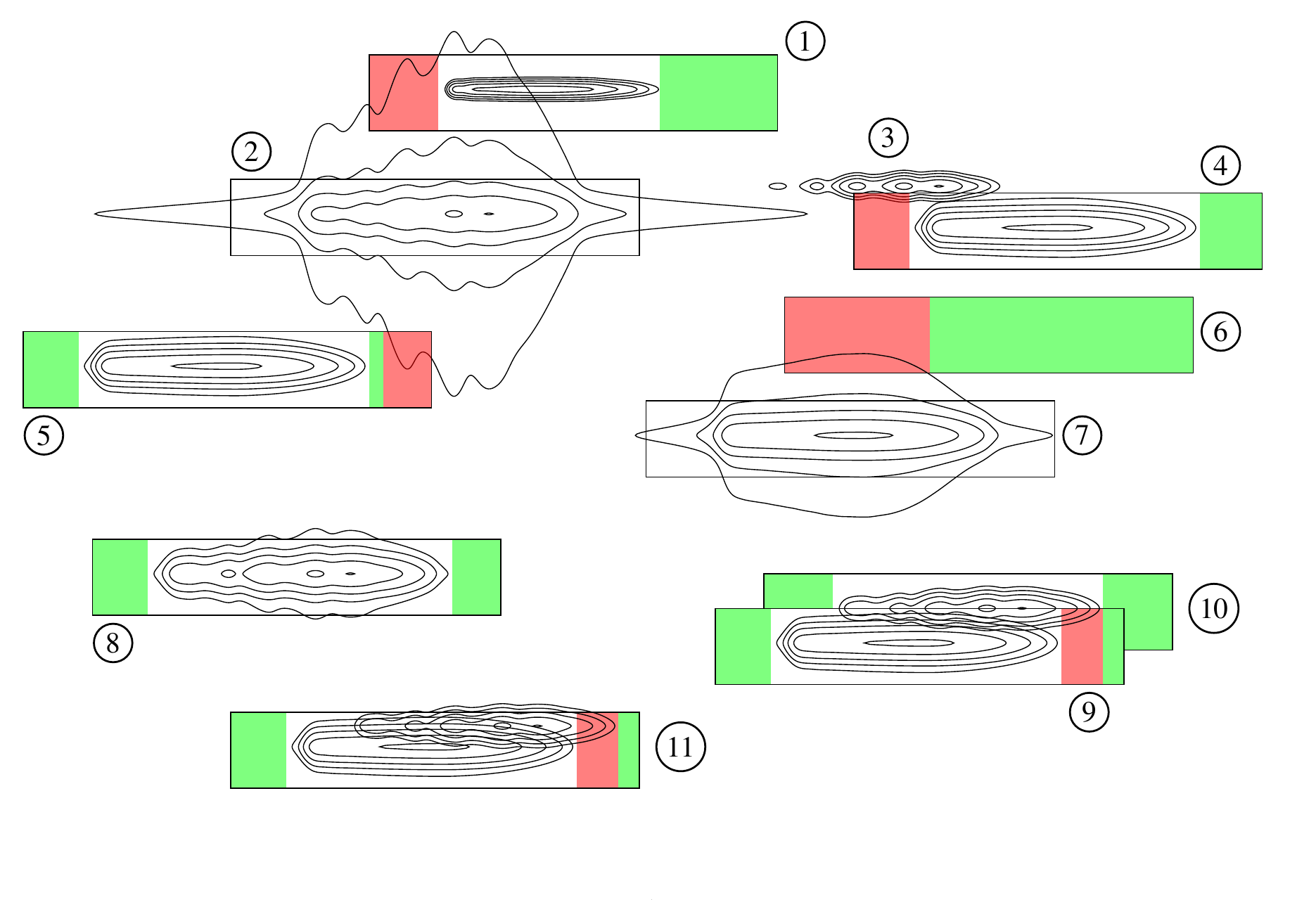}
 \caption{An illustration of the crowding effects on BP and RP dispersed images,
 with their assigned windows (black rectangles). Different cases are
 represented, some going beyond the scope of the present paper. Stars are
 represented by their surface density profiles (solid contours), with the
 outmost contours at the background level. The brightest star (G=15~mag) is in
 window (2), while the faintest one (G=20~mag) is in window (1). Stars fainter
 than the Gaia magnitude limit (G=20.7~mag) are assigned no window, like for
 star (3). The coloured portions of the windows are the background samples, in
 green when they are free from contamination, and in red when they are
 contaminated and cannot be used. When stars are too close, they can be assigned
 to the same widow or two (or more) truncated windows, like in cases (9),
 (10), and (11). There are also empty windows (virtual objects) like in case
 (6). Figure courtesy of A.~Brown.}
  \label{fig:anthony}
\end{figure}

Stars that are further apart than the tiny Gaia PSF are easily deblended by the
PSF fitting algorithms, with results much more similar to HST (Hubble Space
Telescope) than to typical ground-based telescopes. This is the main reason
why -- as we will see in the following -- astrometric measurement uncertainties
(and G magnitudes) are in general less affected by crowding, unlike BP/RP and
RVS measurements.

\subsubsection{BP and RP deblending and decontamination pipelines}
\label{sec:bprp}

\begin{figure}
 \centering
 \includegraphics[width=7cm]{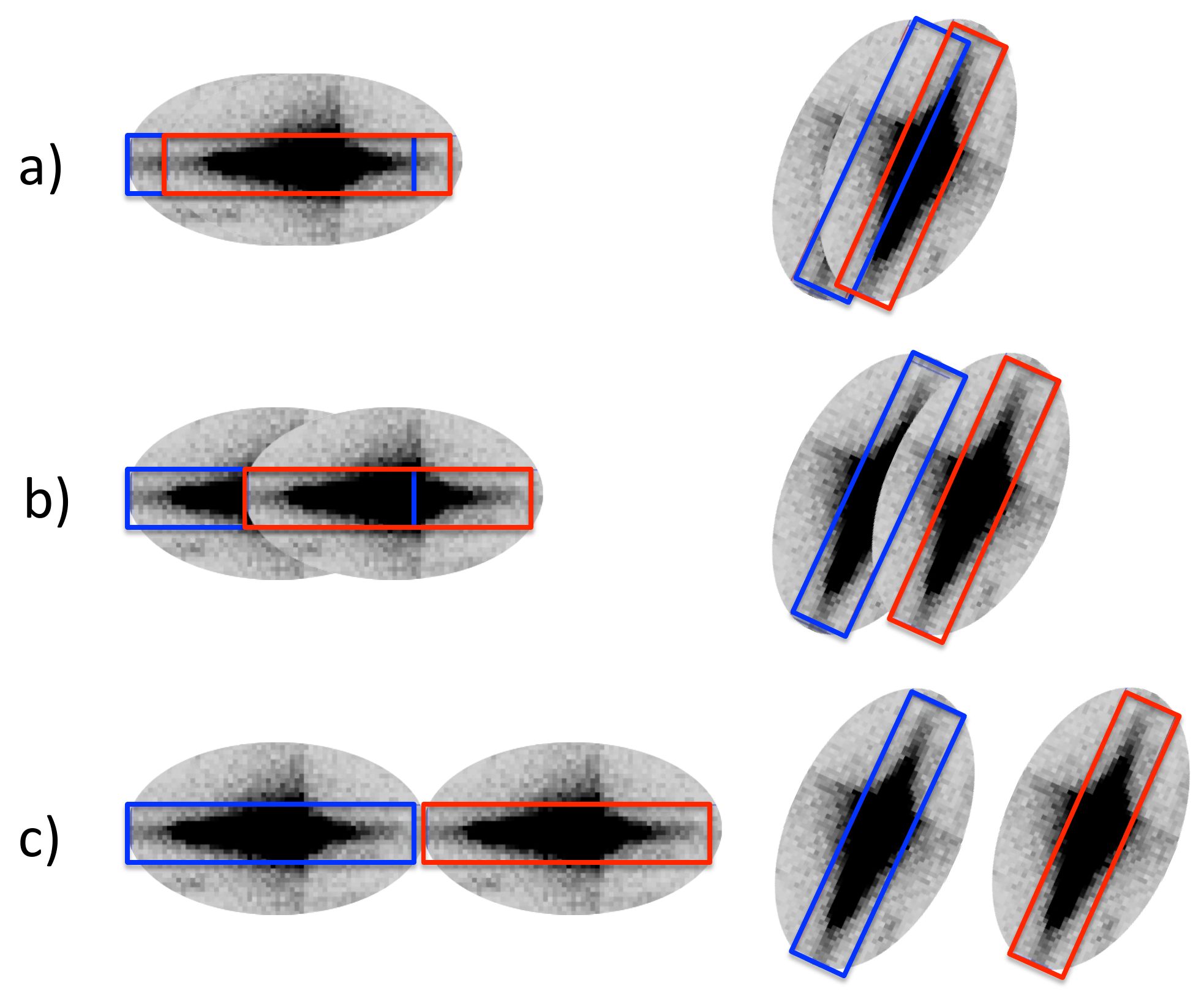}
 \caption{Simplified description of transits in BP/RP. In case (a), two stars
 are closer than half the AC window size, and no matter the orientation of the
 satellite, they will always be assigned the same window (or truncated windows).
 In case (b), two stars have a distance that is in between the AC and AL half
 window sizes, and depending on the orientation, they are sometimes assigned the
 same window (or truncated window), sometimes different windows.  Depending on
 their brightness, they can still contaminate each other's window. In case (c),
 the stars are farther apart than the AL window size, and they are always
 assigned two different windows but, if one of the two stars is very bright, it
 can still contaminate the other significantly.} 
 s\label{fig:classify}
\end{figure}

BP and RP transits of sources that are not isolated will be called here either
{\em blended} or {\em contaminated}. The idea is that when the sources are so
close that they occupy the same window or interfering windows in the vast
majority of the transits ({\em blends}, with D$<$2.12"), they will require a
different treatment than objects that will often be assigned well separated
windows, or transits that are just altered by the flux of a bright source that is
well outside of the window ({\em contaminants}, with D$<$3.54"). This is
illustrated in Figure~\ref{fig:anthony}, where different cases are shown together
with the background evaluation regions, and in Figure~\ref{fig:classify}, showing
how the differing orientation and AL projected distance between sources affects
window assignment.

For blended sources, there are different pipelines that can be used in different
phases of the mission. Blind pipelines (without knowledge of the scene) can be
applied in the initial phases, when the history of each source in different
instruments has not built up to a sufficient level. The two (or more) blended
sources can be roughly modeled without {\em a priori} knowledge of their exact
positions, astrophysical parameters, and fluxes, just by modeling them with two
overlapping spectral energy distributions. This approach was successfully
applied to Gaia commissioning
data\footnote{http://www.cosmos.esa.int/web/gaia/iow\_20150226} of bright stars,
recovering the vast majority of the Ticho-2 binary and double stars. Once a few
transits are accumulated, they can be better disentangled if they are modeled
simultaneously ({\em per source} rather than {\em per transit}), even if there
is still not enough information on the scene, improving the quality of the
reconstruction. Finally, once the scene is well characterized and the history of
the source is well developed, other parameters of the modeling like the spectral
type, the projected distance of the sources along scan, and the relative fluxes,
can be used to further improve the involved sources reconstruction.

For contaminated sources, it is necessary to know well the flux of the
contaminating sources around, thus knowledge of the scene is necessary. Each
known source is modeled to reconstruct the flux even at large distances from the
window (especially for bright stars). The amount of reconstructed contaminating
flux from neighboring sources is computed in each pixel of the contaminated
source window, and subtracted. For all these reasons, decontamination  will have
to be performed contextually with the scene reconstruction and crowding
evaluation.

\subsubsection{RVS deblending and decontamination pipelines}
\label{sec:rvs}

The deblending phylosophy adopted by RVS is slightly different from the one
adopted in AF, or BP and RP. The deblending pipelines are being adapted and
reweitten to mitigate the stray light issues found after
launch\footnote{http://www.cosmos.esa.int/web/gaia/news\_20141217}
\citep{mora16} that impact mostly on RVS spectra, with an expected loss of
$\simeq$1.4~mag in sensitivity \citep{seabroke16}. We therefore based our
modeling of deblending errors on a previous algorithm that was based on general
geometrical considerations \citep{allende08}.

The treatment of contamination is instead part of the background treatment, that
is a vital part of RVS processing, because of the length of the spectra (see
Table~\ref{tab:numbers}). The background is divided in diffused and point-like,
and treated separately. In both cases, a modeling takes place based on knowledge
of the scene and all available data on stray light and nearby objects. The model
produces an estimate of the total background in the RVS windows that is
subtracted from the source signal. Clearly, the model becomes more and more
accurate as Gaia data are progressively accumulated and therefore the best
results will be obtained towards the end of the mission (K.~Janssen,
H.~E.~Huckle, and G.~Seabroke, 2015, private communication).

\begin{figure}
 \centering
 \includegraphics[width=\columnwidth]{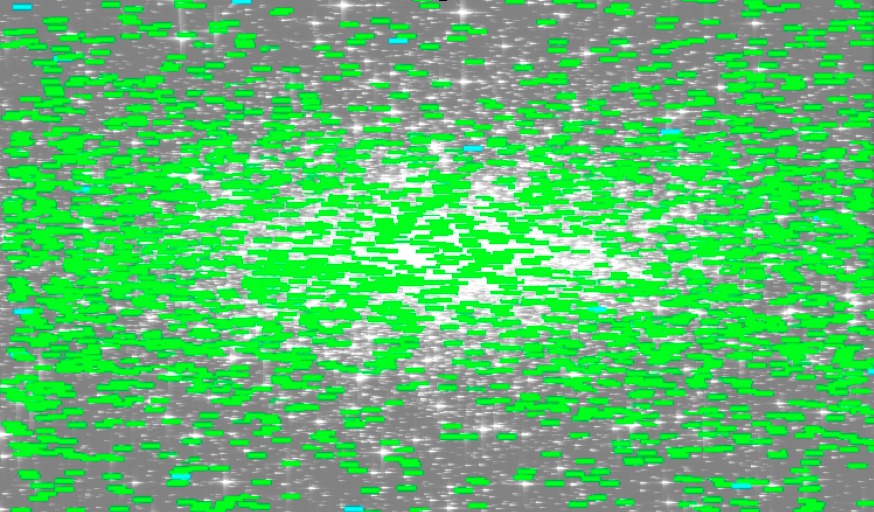}
 \caption{Effect of the on-board detection on completeness. The BP/RP image of a
 GC was simulated with GIBIS \citep[Gaia Instrument and Basic Image
 Simulator,][]{babusiaux05}. The green rectangles are the assigned windows. There
 is an apparent loss of stars above and below the cluster center, where the AC
 stellar density is lower and windows have a lower probabiliy of being assigned
 to a source. The percentage of stars lost is roughly constant with stellar
 density in each transit, but each transit has a different orientation. As a
 results, the end-of-mission incompleteness will tend to be flatter with distance
 from the center than in usual GC photometries. }
  \label{fig:completeness}
\end{figure}

\subsubsection{Image reconstruction and future possibilities}
\label{sec:diana}

To facilitate the study of crowded areas, two paths are being followed. On the
one hand, some crowded regions like the center of $\omega$~Centauri or NGC~1818
in the Magellanic Cloud were imaged and transmitted to the ground in 2D mode in
the SM\footnote{http://www.cosmos.esa.int/web/gaia/iow\_20140206}, during
commissioning, and further regions might be observed in 2D during the mission
lifetime. The goal of these 2D images is to fully reconstruct the {\em scene}
around each Gaia source to help improving and testing the deblending algorithms
in crowded areas. 

On the other hand, pipelines for the 2D image reconstruction from individual
Gaia transits are being developed. One such pipeline, called Source Environment
Analysis (SEA) pipeline, is based on the FastStack image reconstruction
\citep{harrison11,gaia1}. The initial tests are promising and the reconstructed
images could be useful, among other things, to test and improve the deblending
and decontamination pipelines as well. 

Therefore, the simulations presented in this paper, being based on the current
status of the deblending and decontamination pipelines, have to be seen as
generally pessimistic.

\subsection{A note about completeness in Gaia data}
\label{sec:comp}

A particular limitation of the Gaia design is related to the number of
simultaneous object and background (virtual object) windows that can be
allocated by the on-board detection algorithm in each CCD and in any given
moment. This number varies from $\simeq$35\,000 to 1\,050\,000 objects per
square degree, depending on the instrument \citep[see][for more details]{gaia1}.
Areas of the sky that can suffer from this limitation are clearly the Galactic
bulge and plane, and all other fields of view that happen to overlap them on the
Gaia focal plane\footnote{We recall that to obtain absolute astrometric
measurements, Gaia projects two different lines of sight on a common focal
plane. To disentangle sources coming from the two projected lines of sight, each
of the SM (Sky Mapper) CCD columns sees only one of the lines of sight.
Therefore, there will be some fields that -- even with low stellar background --
will happen to overap crowded areas in some fraction of their transits.}. More
relevant to the present study, the limitation applies to the central regions of
GCs, where the local star density can be very high. 

At the single transit level, completeness is also influenced by the on-board
detection algorithm. Every source that enters the SM is assigned a window,
prioritizing brighter objects down to the limiting magnitude. The windows follow
the stars along the focal plane as Gaia scans the sky, and when the stars exit
the field, the window-slots are ``freed" and can be assigned to new sources
again. For this reason, the central dense areas of the clusters will --
statistically speaking -- contain more (bright) stars, and thus obtain more
windows with respect to the periphery of the cluster, in the AC direction.
Different scans, oriented in the sky with different angles, will loose different
stars and thus at the end-of-mission, each star that is not entirely lost will
have lost part of its transits. This is illustrated in the simulated cluster in
Figure~\ref{fig:completeness}, and is also visible in the $\omega$~Centauri
actual Gaia data shown in a press
release\footnote{http://www.cosmos.esa.int/web/gaia/iow\_20141113}.

We do not attempt to simulate these completeness effects in the present paper,
as they would require a full end-of-mission simulation of different lines of
sight in the sky, including the detailed behaviour of the on-board detection
algorithms.

\section{Modeling of crowding errors}
\label{sec:crowderr}

As mentioned above, our goal is to illustrate what Gaia can do for GC studies,
rather than attempting a rigorous simulation of deblending errors. Therefore, we
will model available simulations of crowded Gaia data, to derive simple
formulaes with as few free parameters as possible. 

The crowding error models derivation for different Gaia instruments and
different types of blends are described in the following sections. The models
describe the expected percentage errors on flux caused by crowding as a function
of relevant parameters like the contaminating flux, contaminating colour, and
distance of the contaminating object(s). 

Later, in Section~\ref{sec:ggc_simu} we will transform our flux errors into
errors on other parameters like positions, parallaxes, proper motions, and
radial velocities. We will then use these errors to worsen the nominal Gaia
post-launch science performances, depending on the crowding level suffered from
each of our simulated GC stars. 

\subsection{`Classic' blends}

We term in this paper {\em classic blends} those stars that are closer than the
AF Gaia PSF (0.177", see Table~\ref{tab:numbers}), and that will remain so along
the five years of Gaia observations. Stars that will move apart thanks to their
parallax or proper motion differences, or that have radial velocity differences
that allow to deblend them, will not be considered as classic blends, but
treated as normal Gaia blends (see below). 

\begin{figure}
 \centering
 \includegraphics[width=8.7cm]{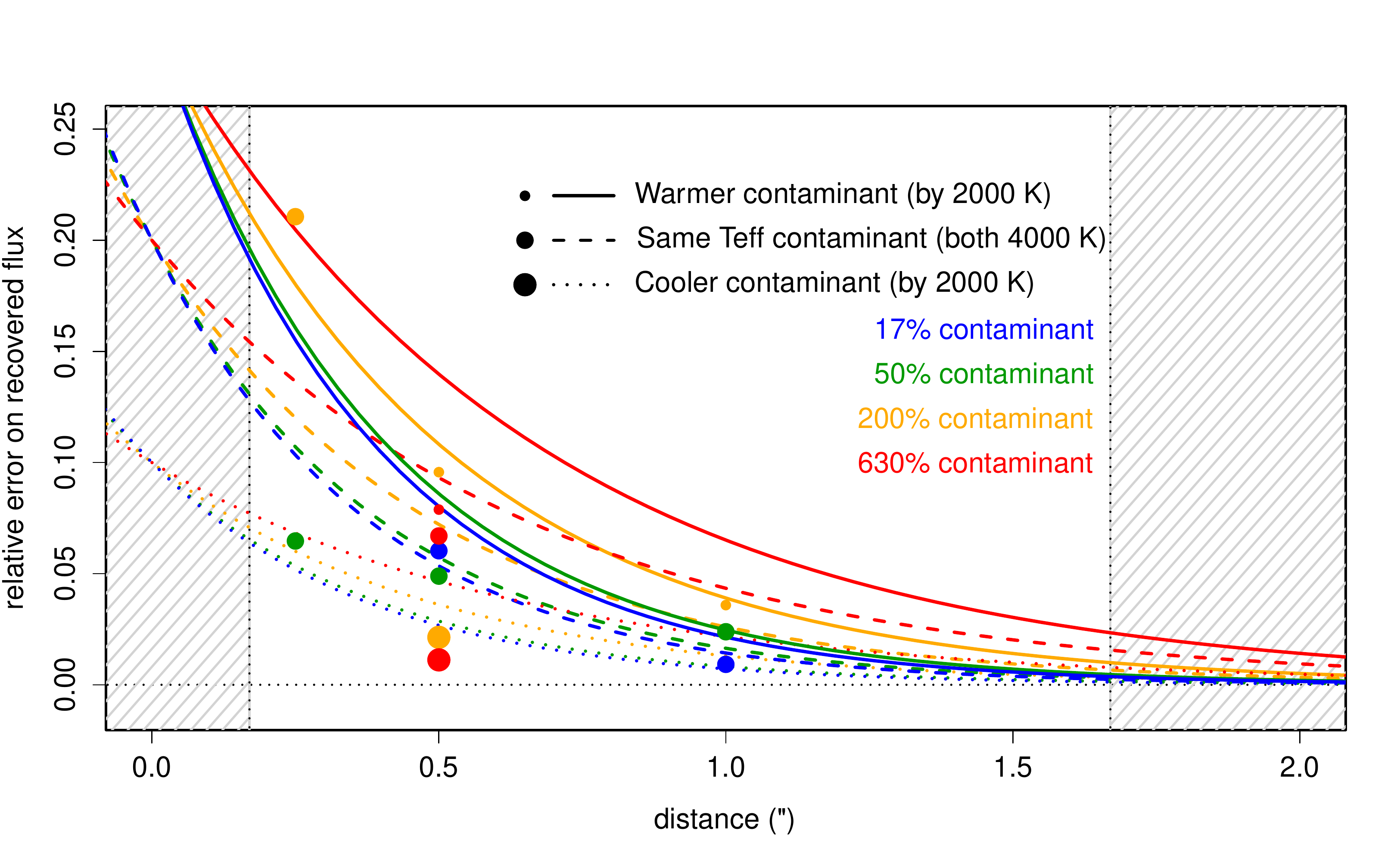}
 \caption{The simulated BP deblending errors as a function of distance (circles),
 based on a G=15~mag star with T$_{\rm{eff}}$=4000~K, and our derived error
 models (curves). The dotted vertical lines mark the applicability range, from
 the Gaia PSF size (0.177") to half the length of the BP/RP windows (1.76").
 Blends with different characteristics are shown: cooler by 2000~K (small circles
 and solid curves), with the same temperature (medium circles and dashed curves),
 and hotter by 2000~K (large circles and dotted curves). The colours refer to the
 relative contaminating flux: 17\% (blue), 50\% (green), 200\% (yellow), and
 630\%(red).}
  \label{fig:XPsimu}
\end{figure}

The precession of Gaia has an important implication for crowding treatment: each
time Gaia scans a particular region of the sky, it is oriented differently, and
thus the projected distance of two stars in the SM columns, on which the
on-board object detection and window assignment are based, are different (see
Figure~\ref{fig:classify}). Therefore, as described in Section~\ref{sec:nss},
even the closest classic blends have a chance to be deblended, although it is
difficult to simulate realistically the errors caused by deblending when so many
observables enter iteratively the deblending procedure along the whole mission. 

In this paper, we will assume that all blends of stars closer than the Gaia PSF
will be recognized as such, thanks to the mission long lifetime, its tiny
astrometric and photometric errors, and sophisticated data analysis methods. 
Because of the limited number of classic blends\footnote{From 5 to 8\% of the
total number of stars in our simulations, including field stars, are classic
blends, depending on the particular cluster. The classic blends are mostly in
the inner parts of GCs, within 1' approximately. For more details see
Section~\ref{sec:crowdeval}.}, this assumption is not going to have a
significant impact on the following analysis. If the relatively small number of
classic blends comes as a surprise, it has to be noted that Gaia observations
only reach V$\simeq$21~mag in GCs, and thus the actual crowding levels are much
lower than those of the typically deeper photometry from the ground or with
HST. 

We will also assume that the actual deblending errors follow the relation adopted
for the BP and RP blends (see below), extrapolating the present simulations
towards smaller source distances. While highly uncertain, this is our only
available estimate of the Gaia deblending capabilities at the present stage. For
RVS there currently are no plans to attempt deblending of sources closer than
0.17", and therefore we will not attempt to simulate the effect of crowding in
these RVS sources.

\subsection{`Gaia' blends}

As discussed in Section~\ref{sec:bprp}, Gaia blends are all those cases in which
two stars are closer than half the AC window size of the relevant instrument
(Table~\ref{tab:numbers}, these all fall into case (a) of
Figure~\ref{fig:classify}). Because the AF image deblending will be simpler (the
AL profile is sharper) than that of BP, RP, and RVS spectra, we will
conservatively base our estimate of crowding errors for deblending on a set of
BP and RVS simulations.

\begin{figure}
 \centering
 \includegraphics[width=8.7cm]{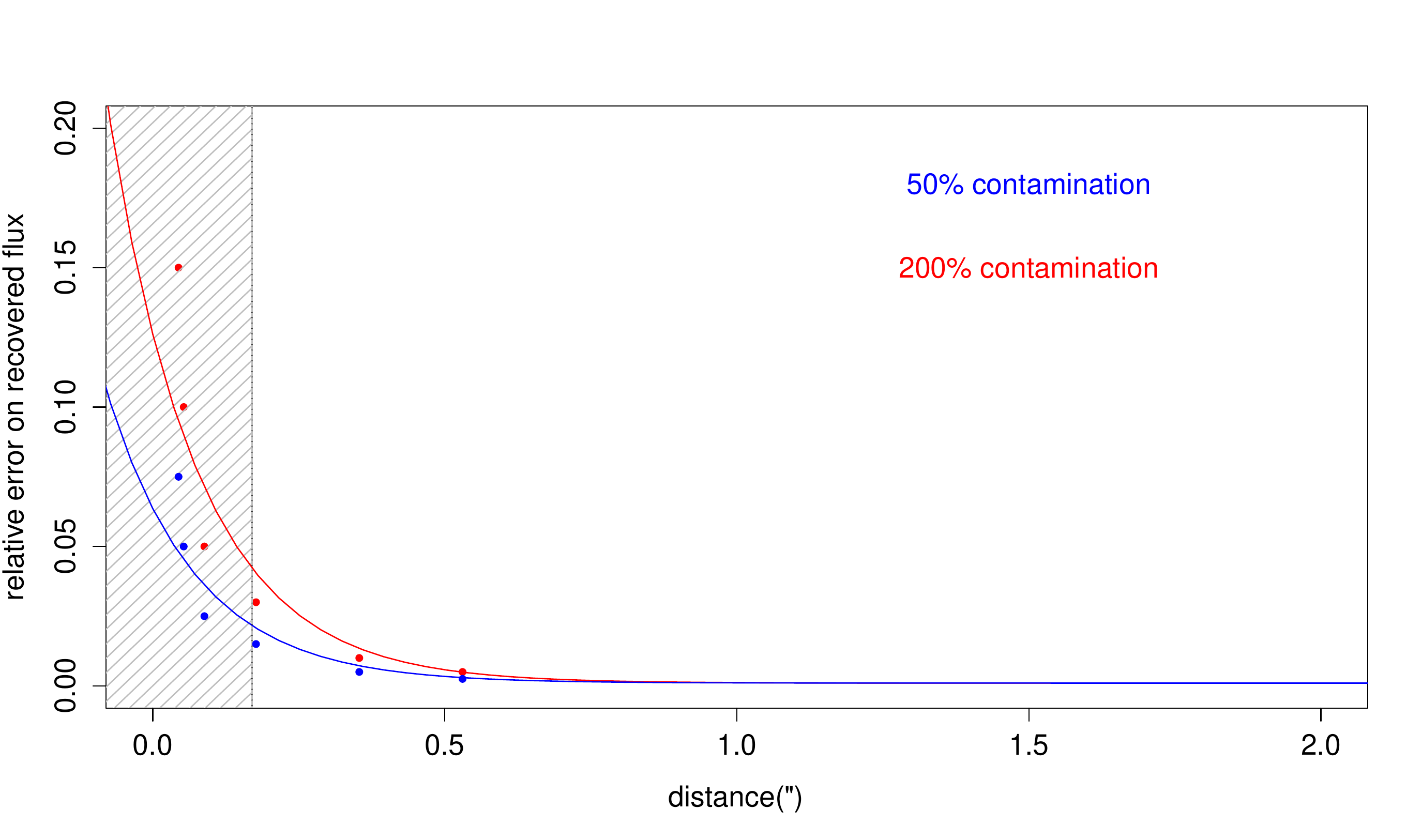}
 \caption{The simulated pre-launch RVS deblending errors, for different
 contamination levels, and our simplified exponential fits. The shaded area
 (D$<$0.17") is not yet treated in the current RVS deblending pipelines, and our
 fits there are extremely uncertain, therefore we will not provide any simulation
 for classic blends in RVS.}
  \label{fig:RVSsimu}
\end{figure}

For the BP and RP blending simulations, we used a per-transit, partially informed
deblending pipeline\footnote{For a detailed description of the various BP/RP
deblending pipelines, see Section~\ref{sec:bprp}. What we mean here by `partially
informed' is that we used information on the blended stars positions in AF, but
we did not use any other information, such as astrophysical parameters
(T$_{\rm{eff}}$, log$g$, A$_V$, [Fe/H]) and therefore we assumed we did not know
the spectral energy distribution of the involved sources. This is still a
pessimistic assumption, because towards the end of the mission this kind of
information will become available and iteratively improved with every data
processing cycle.} on the GIBIS \citep[Gaia Instrument and Basic Image
Simulator,][]{babusiaux05} simulated BP spectra of a typical GC red giant, with
T$_{\rm{eff}}$=4000~K and G=15~mag. The star was blended with other similar
stars, having either T$_{\rm{eff}}$=4000 or 6000~K, G=15.75 or 17.00~mag, and at
different distances equal to 0.25, 0.5, and 1.0$^{\prime\prime}$ (see
Figure~\ref{fig:XPsimu}).

We then modeled the deblending (relative) errors on the recovered flux as an
exponential $\alpha~e^{~\beta\,\rm{D}}$, where D is the distance between the two
blended stars. The coefficients $\alpha$ and $\beta$ vary with the temperature
difference and the relative contaminating flux, respectively, as illustrated in
Figure~\ref{fig:XPsimu}. For RP, the deblending errors were arbitrarily
multiplied by 1.5, owing to the generally wider AL shape of the spectra, that
produces worse deblending results than in BP. These simplified recipes are
intentionally pessimistic, to accommodate for unexpected sources of error.  The
approximate range of validity of the model (highlighted in
Figure~\ref{fig:XPsimu}) goes from the Gaia PSF size to half the AL size of the
BP and RP windows. As discussed in the previous section, we applied an
extrapolation of this relation to D$<$0.17" to classic blends, for lack of a more
detailed study at this stage.

\begin{table*}
\caption{Simulated clusters properties and crowding evaluation results. Fore
each of the 18 simulated GCs we list (see text for more details): (1) the GC ID
number; (2) the GC concentration parameter,
c=log(r$_{\rm{t}}$/r$_{\rm{c}}$); (3) the GC projected distance; (4) the type
of simulated background; (5) the total number of simulated stars (GC and
background) above Gaia detection limit; (6) the number of GC members; (7) the
number of classic blends; (8) the number of Gaia blends; (9) the number of
contaminated stars; (10) the number of clean stars, i.e., contaminated by less
than 0.1\% in flux (roughly 1~mmag); (11) the quick GC designation used in
the paper; and (12) some examples of GCs with similar distance, concentration, 
and background field contamination.
\label{tab:clusters}}
\begin{center}
\begin{tabular}{lccl@{}rrrrrrll}
\hline
\\
Cluster & $c$ & distance & background & $n_{tot}$ &
$n_{GC}$ & $n_{classic}$ & $n_{blends}$ & $n_{contam}$ & $n_{clean}$ &
Designation & Similar to \\
        & (dex)         & (kpc)    &            &&&&& \\
\hline
\\
\# 1  & 1.0 &  5 & halo  &   73385 & 72510 &  4621 & 45101 & 52142 & 16093 & {\em Easy case}         & M13, M92 \\
\# 2  & 1.0 & 10 & halo  &   30200 & 29325 &  3521 & 21550 & 23754 &  4275 &                         & M92 \\
\# 3  & 1.0 & 15 & halo  &   14027 & 13152 &  2203 & 10312 & 11203 &  1495 &                         & NGC~5053 \\
\# 4  & 1.0 &  5 & disk  &   54923 & 26816 &  1028 & 14115 & 17289 &  7379 &                         & M71 \\
\# 5  & 1.0 & 10 & disk  &   33435 &  5328 &   326 &  3302 &  3812 &  1164 & {\em Intermediate case} & M56, NGC~2298\\
\# 6  & 1.0 & 15 & disk  &   29737 &  1630 &   123 &  1048 &  1208 &   311 &                         & M79 \\
\# 7  & 1.0 &  5 & bulge & 1537592 & 71996 &  9833 & 69887 & 71600 &   218 &                         & M22, NGC~6553 \\
\# 8  & 1.0 & 10 & bulge & 1494601 & 29005 &  5620 & 28497 & 28919 &    36 &                         & M9, NGC~6638 \\
\# 9  & 1.0 & 15 & bulge & 1478538 & 12942 &  2998 & 12762 & 12911 &    15 &                         & Pal~11 \\
\# 10 & 2.5 &  5 & halo  &   73385 & 72510 & 11902 & 47043 & 53029 & 15623 &                         & M5 \\
\# 11 & 2.5 & 10 & halo  &   30200 & 29325 &  6914 & 21977 & 23944 &  4117 &                         & M3 \\
\# 12 & 2.5 & 15 & halo  &   14027 & 13152 &  3663 & 10407 & 11176 &  1525 &                         & NGC~5466 \\
\# 13 & 2.5 &  5 & disk  &   54923 & 26816 &  3236 & 15372 & 18085 &  6946 &                         & M92, Pal~10 \\
\# 14 & 2.5 & 10 & disk  &   33435 &  5328 &   906 &  3491 &  3968 &  1055 &                         & M79, NGC~1851 \\
\# 15 & 2.5 & 15 & disk  &   29737 &  1630 &   292 &  1097 &  1250 &   300 &                         & M15 \\
\# 16 & 2.5 &  5 & bulge & 1537592 & 71996 & 16564 & 70125 & 71648 &   190 &                         & NGC~6540, NGC~6558 \\
\# 17 & 2.5 & 10 & bulge & 1494601 & 29005 &  8726 & 28556 & 28924 &    40 &                         & NGC~6325, NGC~6342 \\
\# 18 & 2.5 & 15 & bulge & 1478538 & 12942 &  4378 & 12770 & 12914 &    13 & {\em Difficult case}    & M54, NGC~6517 \\
\hline
\end{tabular}
\end{center}
\end{table*}

For RVS, the only available deblending simulations were computed before launch
\citep[][see also Figure~\ref{fig:RVSsimu}]{allende08}, when the windowing scheme
and backgound treatment were different (see Section~\ref{sec:rvs}). Thus, the
error modeling presented here will be much more uncertain for RVS than for BP and
RP, or for AF. The temperature or colour difference between two sources has little
impact on the deblending ability of the RVS pipelines, because the spectral shape
is approximately flat along the RVS windows. What mostly counts is the relative
contaminating flux. As a result, the best-fitting exponential laws had a constant
$\beta$, and $\alpha$ varying with the relative contaminating flux. As mentioned
above, we did not extend the model below the PSF size for RVS (shaded region in
Figure~\ref{fig:RVSsimu}).

\subsection{Contaminants}
\label{sec:contam}

For contaminated stars (cases (b) and (c) in Figure~\ref{fig:classify}), the
reconstruction of star fluxes at distances larger than the window size is
subject to a variety of uncertainty sources, and the relevant pipelines will be
effective only when a sufficient history is accumulated to build a scene. Not
only the stars positions are needed, but also their spectral energy
distributions. Additionally, a complete characterization of the PSF profiles AC
and AL, at different wavelengths and positions on the focal plane will be
fundamental. Therefore no detailed simulations are available at the moment
(Section~\ref{sec:bprp}). 

However, we do know that a step-like behaviour is expected in the errors at
1.76$^{\prime\prime}$, where the processing switches from decontamination to
deblending, in the sense that decontamination pipelines perform slightly worse
than the deblending ones. There will be, on purpose,  a `grey area' in projected
distance between sources, where both pipelines will be applied, for
cross-validation purposes. Some tests were performed to develop the current
decontamination algorithms \citep[][, and De Luise, private
communication]{piersimoni11}. According to those tests, we decided to roughly
approximate the decontamination errors with a factor of 2 worsening with respect
to deblending errors. This is a pessimistic assumption.

Our modeling of decontamination residual errors is applied to contaminant stars
with distances ranging from half the AC window size to the full AL window size or
to the contaminating star's `size', whichever is larger. We modeled the
contaminating stars sizes as a linear function of G magnitude (L.~Pulone and
P.~M.~Marrese, 2008, private communication, see also Figure~\ref{fig:anthony}).
Stars fainter than G$\simeq$15~mag are generally smaller than the BP and RP AL
window size\footnote{With our linear relation, a G=2~mag star would be able to
significantly contaminate stars up to 34.4$^{\prime\prime}$. These bright stars
are rare: in our eighteen simulated GCs there is only one field star brighter
than that, appearing in the six disk GCs.}.

For RVS, the discrete background pipelines are being presently integrated and
tested, so no simulations are available at the moment and we just extended our
simplified treatment of the RVS blends (previous section) to RVS contaminants,
as done for BP and RP, using the appropriate window sizes.

\begin{figure}
 \centering
 \includegraphics[width=\columnwidth]{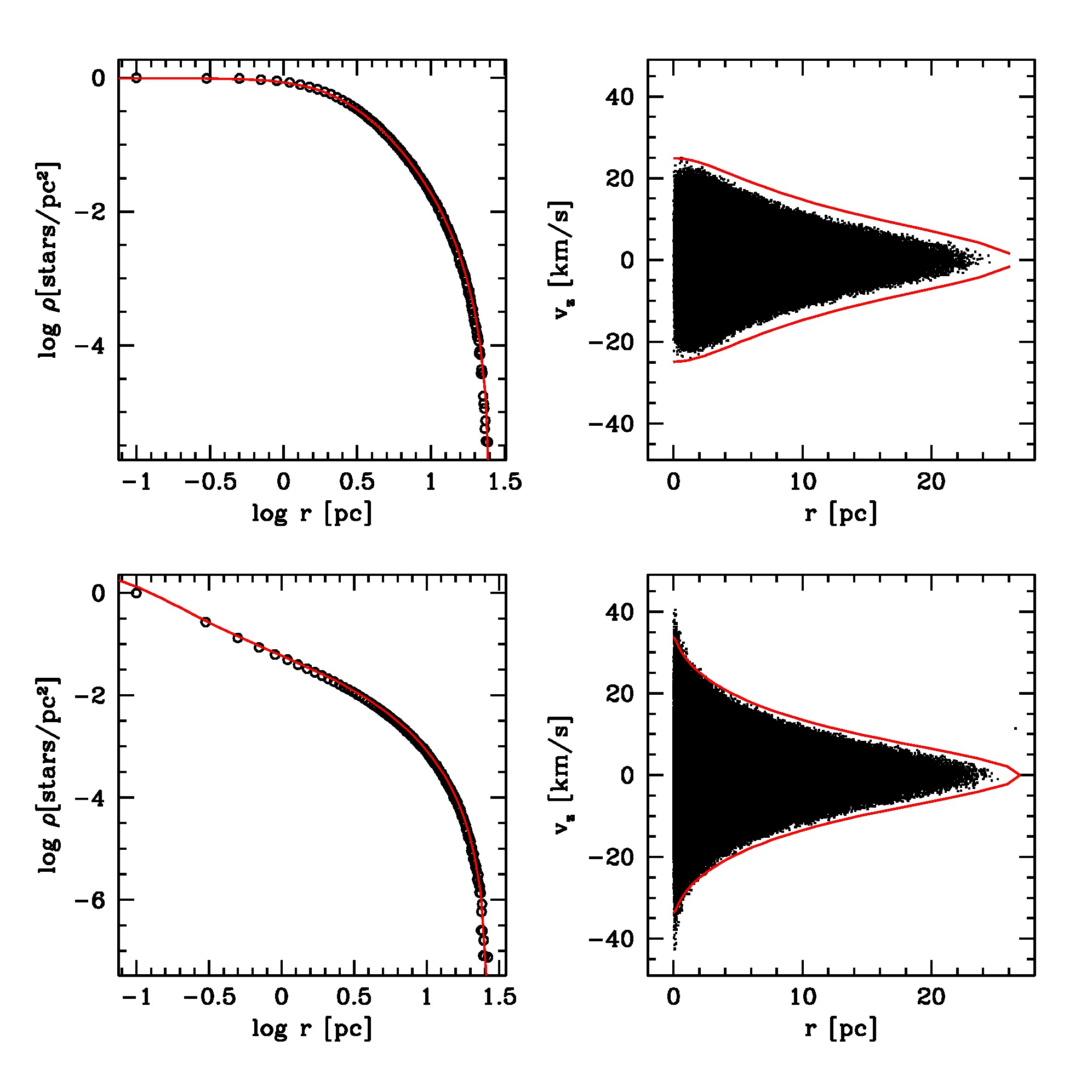}
 \caption{The logarithmic density profiles (left panels) and radial velocity
 dispersion profiles (right panels) for the two synthetic clusters with
 concentration parameter c=1.0 (top panels) and c=2.5 (bottom panels). The red
 lines are the corresponding theoretical density profiles (left panels) and
 $\pm$3\,$\sigma$ contours about the systemic velocity for \citet{king66} models
 with the same parameters.}
  \label{fig:clusters}
\end{figure}

\section{Cluster simulations}
\label{sec:ggc_simu}

We simulated 18 different globular clusters, as summarized in
Table~\ref{tab:clusters}, with different concentration, background contamination,
and distance. The final simulated clusters, after the Gaia science performances
and the crowding error simulations are included, are presented in
Table~\ref{tab:cat}.

\subsection{3D simulated clusters}
\label{sec:3Dsimu}

The simulated clusters were computed with the McLuster
code\footnote{http://www.astro.uni-bonn.de/$\sim$akuepper/mcluster/mcluster.html}
\citep{kupper11}, that is designed to produce initial conditions for N-body
simulations. We used the code version ({\em mcluster\_sse}) which implements the
stellar evolution recipes by \citet{hurley00}.

We produced two clusters containing 800\,000 stars, with positions and velocities
drawn from equilibrium \citet{king66} models. The two clusters differed only in
their concentration parameter, c=log(r$_{\rm{t}}$/r$_{\rm{c}}$), one having c=1.0
and the other c=2.5. Among the key input parameters for the simulation, the
half-mass radius was chosen to produce a (projected) half-light final radius
similar to what typically observed for Galactic globular clusters: 4~pc for the
c=1.0 cluster and 3~pc for c=2.5 one \citep[see also][and following
updates]{harris96}. Among the key input stellar population parameters, we adopted
an age of 12~Gyr, a metallicity of Z=0.0003, corresponding to [Fe/H]=--1.79~dex,
and a \citet{kroupa01} IMF (Initial Mass Function). The simulations were
performed without including binaries, for simplicity, and the resulting clusters
are spherical and non-rotating.

The main implications of our choices of input parameters are: {\em (i)} a blue
HB (Horizontal Branch) morphology, which may be useful to test the performance
of Gaia on relatively hot stars, and {\em (ii)} a stellar M/L$_{\rm{V}}$ ratio
larger ($\geq$3 instead of $\leq$2) than what typically observed in Galactic
globular clusters having a present day mass function compatible with a
low-exponent power law \citep{paust10}. The high M/L$_{\rm{V}}$ ratio is also
due to a relatively large fraction of dark remnants: $\simeq$25\% of the cluster
mass is contributed by objects with M$>$M$_{\odot}$\footnote{Dark remnants are
white dwarf stars, neutron stars, and black holes. Recently, comparable
fractions of dark remnants were obtained with state-of-the-art cluster modeling
by \citet{sollima15} and \citet{sollima16}.}. Independently of the recipes
giving raise to such a large fraction of dark remnants and their actual nature,
this is not a concern for our purpose: the only effect is that the potential
well may be slightly deeper than for most real clusters and, consequently, the
velocity dispersion is slightly larger, at fixed total luminosity and scale
radius. What is relevant for us is that the central velocity dispersion is still
in the right range and that the model is self-consistent.

The resulting simulated clusters have absolute integrated magnitudes of
M$_{\rm{V}}$=--7.6~mag, i.e., just above the peak of the clusters luminosity
function \citep[M$_{\rm{V}}$=--7.4~mag, according to][]{brodie06}. The clusters
projected surface density distributions are well fitted by a \citet{king66}
profile, when obtaining r$_{\rm{c}}$ (the core radius) from the fit. Finally,
the derived central velocity dispersions agree well with the predictions by
\citet{king66}, where the total mass is obtained by summing the masses of all
individual stars. 

\subsection{Simulated field contamination}

We simulated the field Galactic population in three directions using the
Besanc\c on models\footnote{http://model.obs-besancon.fr/js} \citep[][and
references therein]{robin03}, including kinematics and without observational
errors. The models produced catalogues of magnitudes (in the Johnson-Cousins
system), 3D positions, and 3D motions. We adopted the ``standard"
extinction law offered by the Besanc\c on simulator, a mean and diffused
absorption of 0.7~mag/kpc, neglecting small scale variations, and decreasing
away from the Galactic plane with a smooth Einasto profile \citep[for more
details, see][]{robin03}. We selected all available spectral types in a
magnitude range of 0$<$V$<$25~mag. The simulations were computed  in a
0.7$\times$0.7~deg square centred on each of the following three directions (see
also Table~\ref{tab:clusters}):

\begin{itemize}
\item{a rather empty {\em ``halo"} field at l=150 and b=80~deg, containing
$\simeq$2300 stars in total; this is meant to represent the best cases of low
background contamination in Gaia observations;}
\item{an extremely dense {\em ``bulge"} field at l=5 and b=5~deg,
containing almost 6 millions of stars, which should cover even the most crowded
Gaia observations, when two relatively crowded lines of sight overlap on the
focal plane (see also Section~\ref{sec:gaia});}
\item{a crowded {\em ``disk"} field in the anticenter direction, at l=180 and
b=0, containing roughly 140\,000 stars, with a relatively high extinction of up
to $\simeq$2~mags in V band \citep[even if these values are only reached in
$\simeq$25\% of the known GCs in][]{harris96}; while not as extreme as the bulge
line of sight, this direction provides still severe crowding levels -- on the
Galactic plane -- coupled with significant reddening.}
\end{itemize}

\begin{table}
\caption{Column-by-column description of the final simulated stars, that will
only be available in the electronic edition of the journal, and at CDS.
\label{tab:cat}}            
\begin{center}         
\begin{tabular}{lrcl}     
\hline
\\ 
Content                 & Column & Units & Description \\
\hline
\\
Cluster                 & (1)    &                     & Cluster number \\
Star                    & (2)    &                     & Star ID      \\
RA$^\prime$             & (3)    & (deg)               & Position along the RA direction \\
Dec$^\prime$            & (4)    & (deg)               & Position along the Dec direction \\
$\delta$coord           & (5)    & (deg)               & Error on RA$^\prime$ and Dec$^\prime$ \\
$\mu_{\rm{RA^\prime}}$  & (6)    & (mas~yr$^{-1}$)     & RA$^\prime$ proper motion \\
$\mu_{\rm{Dec^\prime}}$ & (7)    & (mas~yr$^{-1}$)     & Dec$^\prime$ proper motion \\
$\delta \mu$            & (8)    & (mas~yr$^{-1}$)     & Error on proper motion \\
$\varpi$                & (9)    & (mas)               & Parallax \\
$\delta \varpi$         & (10)   & (mas)               & Parallax error \\
G                       & (11)   & (mag)               & G-band integrated magnitude \\
$\delta$G               & (12)   & (mag)               & G magnitude error \\
G$_{\rm{BP}}$           & (13)   & (mag)               & G$_{\rm{BP}}$ integrated magnitude \\
$\delta$G$_{\rm{BP}}$   & (14)   & (mag)               & G$_{\rm{BP}}$ magnitude error \\
G$_{\rm{RP}}$           & (15)   & (mag)               & G$_{\rm{RP}}$ integrated magnitude \\
$\delta$G$_{\rm{RP}}$   & (16)   & (mag)               & G$_{\rm{RP}}$ magnitude error \\
Membership              & (17)   &                     & True membership \\
\hline                                               
\end{tabular}                                         
\end{center}
\raggedright{Note: The pseudo coordinates RA$^\prime$ and Dec$^\prime$ are just
distances from the center in deg, not true sky coordinates. The true membership, 
in the case of classic blends, relates to the brightest star of the blend. It is 
1 for cluster members and 0 for field stars.}
\end{table}

\begin{figure}
 \includegraphics[width=\columnwidth]{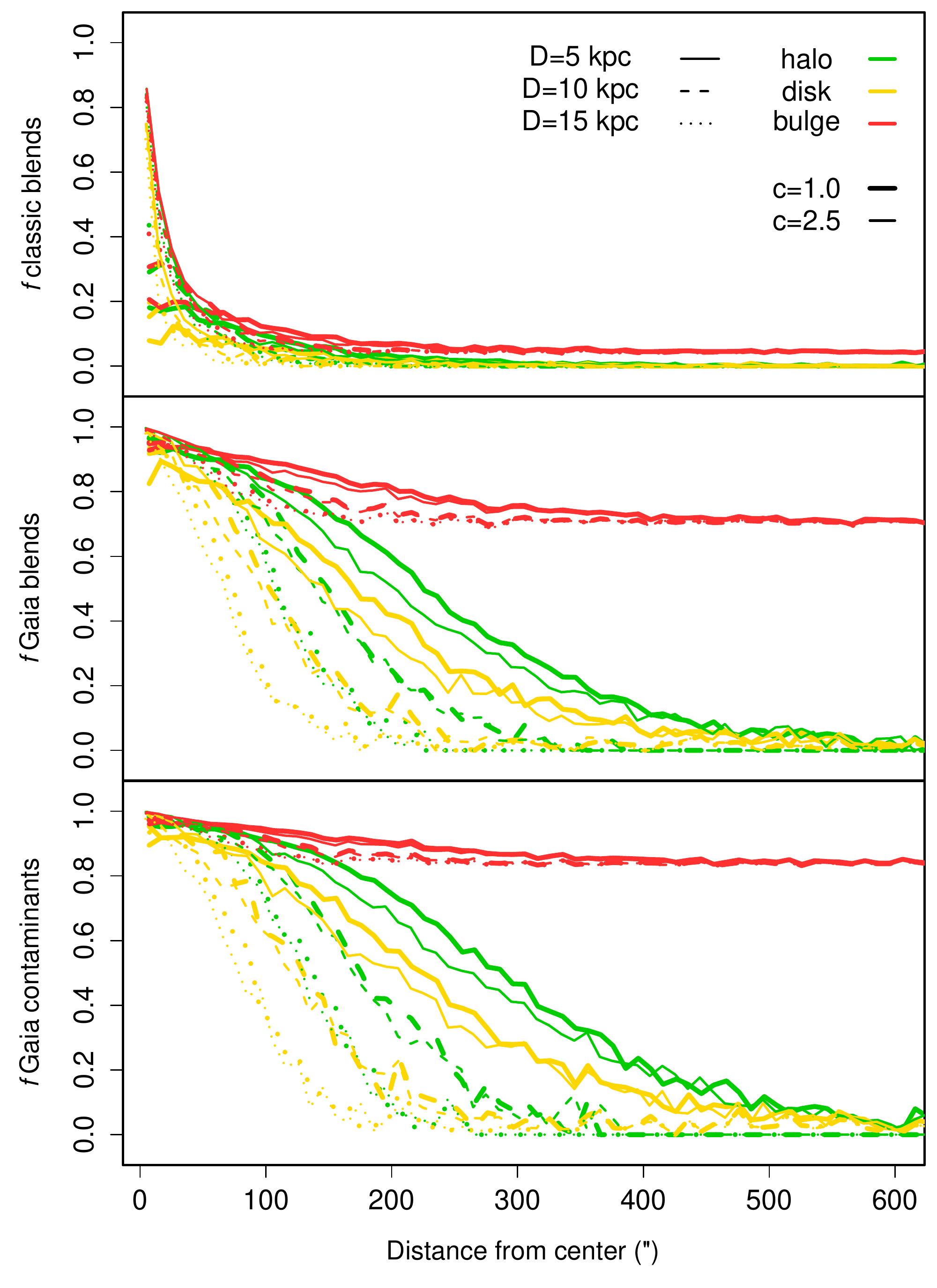}
 \caption{Results of the crowding evaluation on simulated GCs. Each panel
 displays the relative fraction of classic blends, Gaia blends, and
 contaminants as a function of distance from the GC center. Only stars that are
 blended or contaminated by at least 0.01~mags in total (i.e., $\simeq$1\% of
 their flux) are shown in the figure. Only classic blends have a significant
 effect on the astrometric performances, while BP/RP and RVS are significantly
 affected also by Gaia blends and contaminants.}
  \label{fig:counts}
\end{figure}

\subsection{Simulated Gaia measurements}

Each of the two clusters (see Section~\ref{sec:3Dsimu}) was projected at three
different distances of 5, 10, and 15 kpc and combined with each of the three
backgrounds, to produce the 18 clusters in Table~\ref{tab:clusters}. A typical
systemic radial velocity of 100~km~s$^{-1}$ was assigned to each cluster
\citep[broadly compatible with the values listed in the][catalogue, in its most
recent version]{harris96}, while a proper motion of --5~mas~yr$^{-1}$ in both
the RA and Dec was assigned to all clusters \citep[broadly compatible with
the values measured by, e.g.,][]{dinescu99}. Distances were
also converted into parallaxes.

Integrated Gaia magnitudes G, G$_{\rm{BP}}$, and G$_{\rm{RP}}$ (see also
Section~\ref{sec:gaia} and Figure~\ref{fig:filters}) were obtained for each star
using the formulae provided by \citet[][using their coefficients from
Table~3]{jordi10}, starting from the B--V colour. For extinction, we adopted the
pessimistic hypothesis that the cluster reddening was equal to the corresponding
background field's highest value (i.e., as if the cluster was entirely behind
the field population). While this assumption is not completely realistic, it
gave us the possibility of exploring higher reddening conditions, at least for
the disk projections. This yielded an A$_{\rm{V}}$ of 0.062~mag for the halo
projections, 0.080~mag for the bulge projections, and of 2.149~mag for the disk
projections. 

\subsection{Crowding evaluation}
\label{sec:crowdeval}

For each simulated star in each GC, we counted all neighbors (with flux above
0.1\% of the star) within a specified distance, depending on the considered
instrument and on the types of blends (Section~\ref{sec:crowderr}). We adopted
an exhaustive algorithm for neighbors search\footnote{In other words, we looped
over the list of (relevant) stars and computed distances with all other
(relevant) stars, one object at a time. This was necessary because smarter
neighbour searches pre-compute a distance matrix to increase computation speed,
and therefore tend to saturate computer memory for datasets containing more than
10$^5$ objects \citep{hendra}. Our bulge simulated GCs contain a few million
stars, and several tens of millions of relevant pairs.}. 

The relative number of each type of neighbours (classic blends, Gaia blends,
contaminants) for the 18 simulated clusters is shown in Figure~\ref{fig:counts}
as a function of distance from the GC center. As can be seen, the parameter that
dominates the fraction of classic blends is the GC concentration: core collapsed
clusters have more than twice the number of classically blended stars in their
central parts compared to normal GCs. The vast majority of classic blends
are in the GC central regions, and are blends of GC members with other GC
members. Distance and reddening have a curious effect: because GCs become
fainter with increasing distance and reddening, their luminosity function
crosses the detection limit of Gaia where it is less populous. As a consequence,
not only the number of stars detected by Gaia decreases, but also the
probability of blending, and so the blends and contaminant fractions. On the
other hand, the effect of field contaminants is to increase the fraction of
Gaia blends and contaminants at all distances from the center. Extreme cases
are the bulge background GCs, where the vast majority of stars are blended
and/or contaminated to some degree at all radii, mostly by field stars.

We listed in Table~\ref{tab:clusters} some key quantities resulting from the
crowding evaluation. As can be seen, the number of stars that are clean (i.e.,
contaminated by less than 1~mmag) can be substantial when the background is not
extreme. The vast majority of Gaia blends and contaminants will be known, as a
large fraction of the classic blends (see  Section~\ref{sec:nss}). Therefore, it
will always be possible to select a reliable sample of clean stars from Gaia
data. 

Additionally, we report here that already in the first Gaia release, which does
not contain stars in truncated windows or stars with low-quality measurements
\citep{gaia2}, more than 200\,000 stars were actually detected in the
$\omega$~Cen field. This compares well with our \#1 cluster, which is reasonable
considering the higher mass of $\omega$~Cen, and gives support to the figures
discussed above and reported in Table~\ref{tab:clusters}.

\subsection{Simulated Gaia errors}
\label{sec:errors}

To compute the Gaia errors, we combined the post-launch Gaia science
performances \citep{gaia1} with our crowding error estimates. For the magnitudes
we combined the two errors in quadrature, because they are fully independent.
For the astrometric and spectroscopic measurements, the crowding errors are not
fully independent from the Gaia science performances, because they both depend
on the star magnitude. Therefore we conservatively summed the errors in
modulus.

\subsubsection{Crowding errors}
\label{sec:crowd2}

We used the modeling described in Section~\ref{sec:crowderr}, which provided a
relative error on the flux of each star, caused by the various type of blends and
contaminants affecting it. These modeled errors represent well the crowding
errors for G, G$_{\rm{BP}}$, and G$_{\rm{RP}}$, i.e., on the photometry. 

To evaluate the crowding errors on astrometry, we considered two components: the
error on centroiding and a chromatic shift. For the centroiding errors, only
classic blends were considered, because only stars closer than the PSF are
expected to have a significant impact on the final centroid measurement of a
star. We expect that the PSF of a star will be perturbed by the residual flux
left by the close companions after the deblending procedure, and these residuals
will depend on the contaminating flux and distance of the blending stars, as
modeled in Section~\ref{sec:crowderr}. We therefore computed the percentual flux
in the residuals as an average of the bended star and the blending ones,
weighted on distance and flux of the blending stars. We then assumed that the
percentual centroiding error caused by blending was equal to this residual
contaminating flux, a pessimistic assumption. The resulting centroiding errors
caused by crowding are typically of the order of $\sim$10~$\mu$as~yr$^{-1}$
(with maximum values of $\sim$100), but they affect a relatively low number of
stars in each GC (see Table~\ref{tab:clusters}).

For the chromaticity correction, we know that chromatic effects for Gaia are
relevant, because of its high accuracy, and they are corrected using the BP/RP
information. If an error is made on that colour determination because of blends,
the error is transported into the astrometric measurements as well. It was
estimated that the shift on the single Gaia measurement could amout to
500--800~$\mu$as, depending on the spectral type and on the position along the
focal plane. By averaging out transits and by correcting the effect of spectral
type with G$_{\rm{BP}}$-G$_{\rm{RP}}$, the residual end-of-mission errors would
amount to 0.4--1.4~$\mu$as \citep{jordi06}\footnote{These estimates were
computed using the Gaia filter system, now abondoned, but there are no other
estimates available of the Gaia residual chromatic displacement in the
literature at the moment.} in most of the well-behaved stars. However, for
objects with non-stellar or peculiar colours, such as a Quasar or a (multiple)
stellar blend, the residual error could be still be roughly 30~$\mu$as, as
computed from the typical residual centroid shift difference between a B and an
M star, after chromaticity correction \citep{jordi06}. We thus approximated
linearly\footnote{The entity of the chromatic correction errors caused by
crowding is small for the simulated GCs (a few $\mu$as or $\mu$as~yr$^{-1}$)
compared to the Gaia science performances (see also Figure~\ref{fig:epm}), and
therefore a linear approximation was considered adequate.} the additional
chromatic error as 0.002~$\mu$as for each K of temperature difference between
each blended stars pair (both classic and Gaia blends), multiplied by the
relative contaminating flux. The typical residuals from chromatic errors caused
by crowding are of a few $\mu$as, but can be substantially higher for companions
that are very bright or have a very different colour (up to
$\simeq$30--50~$\mu$as in our simulations).

For RV error computations, the first consideration is that everytime a star is
contaminated by another star with a net RV difference above
$\simeq$30~km~s$^{-1}$ (roughly the Gaia resolution element FWHM), there is a
chance of separating the two RVs, depending on the relative flux contamination.
The percentage of flux contamination (obtained as in Section~\ref{sec:crowderr})
was used to degrade the signal of the contaminated (or blended) star. We thus
recomputed the magnitude of each blended or contaminated star, taking into
account the residual errors after deblending or decontamination. We similarly
recomputed the colours taking into account the residual colour errors. The
recomputed colours and magnitudes were fed into the RV error equation
\citep{gaia1}, to derive the error on RV implied by the magnitude and colour
deblending errors. The second consideration is that an artificial RV shift of
roughly 30~km~s$^{-1}$ can also be obtained if two blended stars are offset, in
the AL direction, by at least 4 AL pixels (or 0.235"), therefore we treated
these stars as in the above case. We finally did not attempt to simulate any RV
measurement for stars closer than the Gaia PSF (i.e., one AC pixel or 0.177") as
explained previously. Moreover, we did not compute RVs or RV errors for stars
fainter than G=16~mag.

\begin{figure}
 \centering
 \includegraphics[width=0.95\columnwidth]{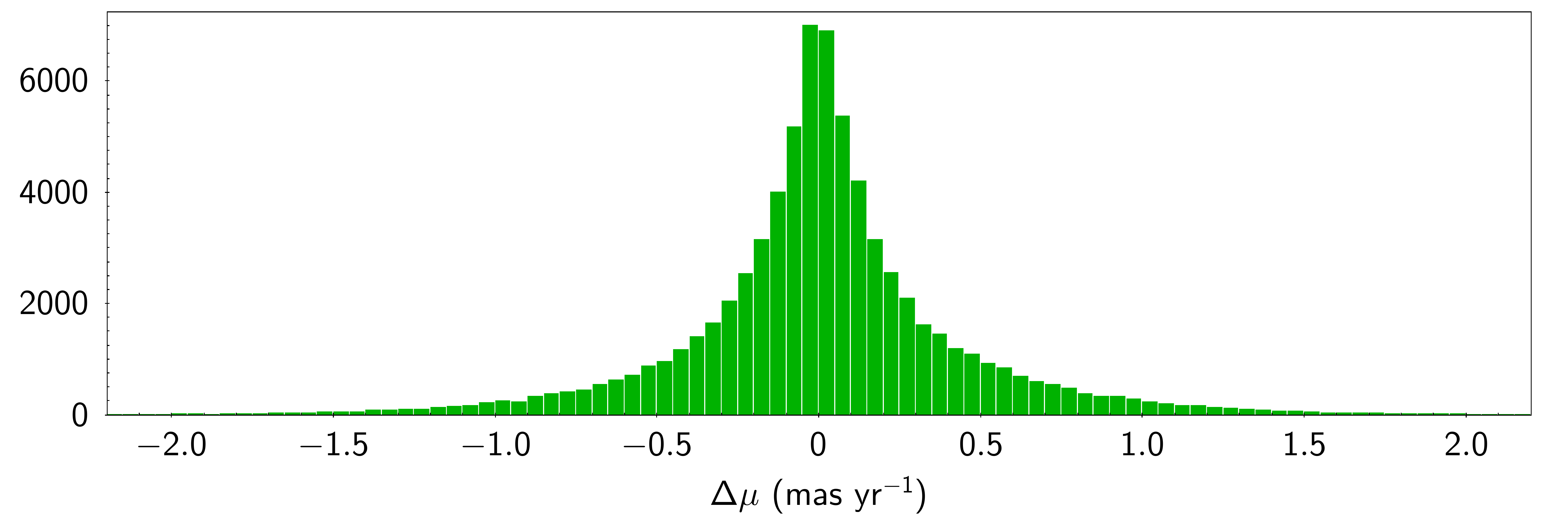}
 \includegraphics[width=0.95\columnwidth]{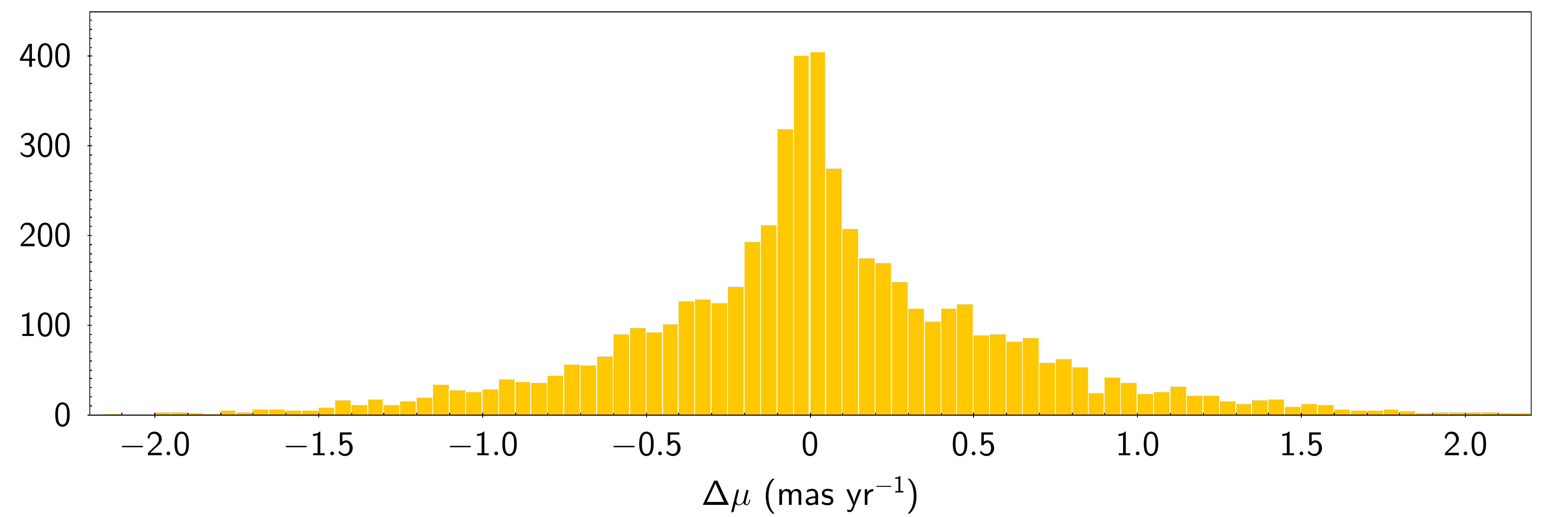}
 \includegraphics[width=0.95\columnwidth]{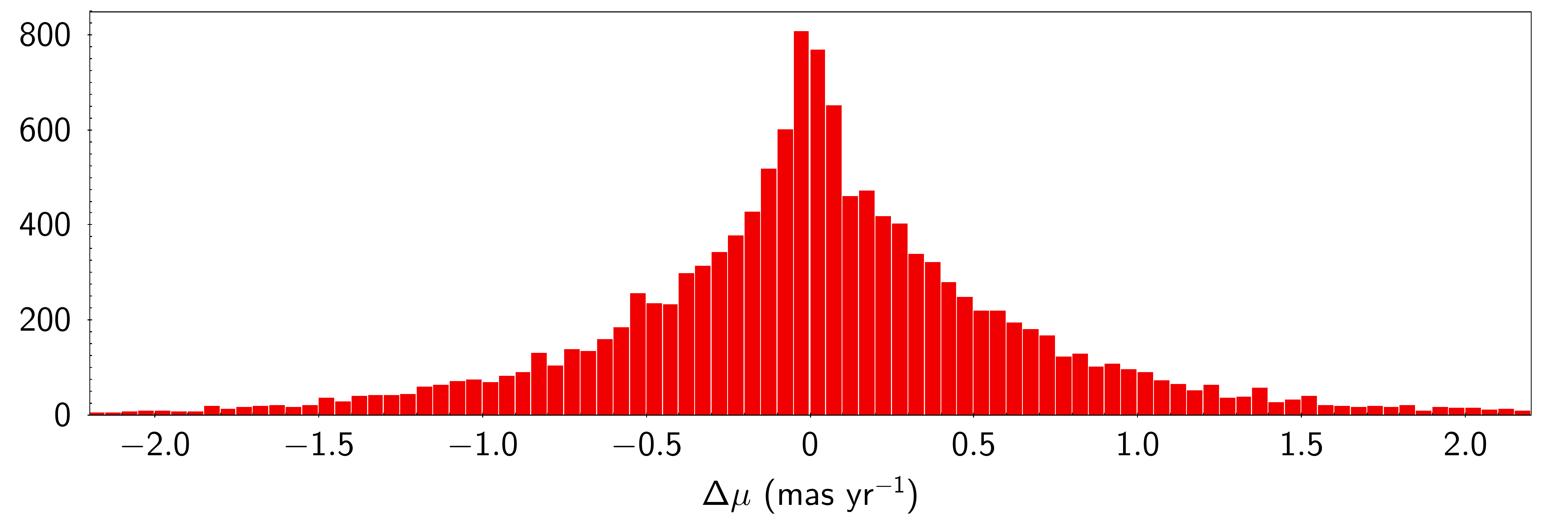}
 \caption{Histograms of differences between the final and error-free simulated
 proper motions of the easy (green histogram), intermediate (yellow histogram),
 and difficult cases (red histogram). Only cluster members are shown. The
 resulting absolute displacements in the systemic proper motion determination is
 of about 1--10~$\mu$as~yr$^{-1}$, corresponding to errors of $<$0.1\%. We note
 that the intrinsic dispersion is of the order of 100~$\mu$as~yr$^{-1}$,
 and the final, simulated one is 300--600~$\mu$as~yr$^{-1}$, when including all
 stars down to G$\simeq$20.7~mag.}
  \label{fig:hpm}
\end{figure}

\section{Results}
\label{sec:res}

We illustrate in the following sections the type of data quality we expect from
Gaia at the end of the mission, based on our simulations. As mentioned already,
given all the conservative assumptions and the use of preliminary deblending
pipelines, the presented results have to be considered pessimistic, in the sense
that the pipelines will be more sophisticated in a few years from now. 

We will show in all figures three simulated GCs: an {\em easy case} with
D=5~kpc, c=1.0, and a halo background (cluster \#1, plotted in green in all
figures); an {\em intermediate case} with 10~kpc, c=1.0, and a disk background,
with a relatively high reddening for GCs (cluster \#5, plotted in yellow); and a
{\em difficult case} with D=15~kpc, c=2.5, and a bulge background, which is an
extreme condition for Gaia (cluster \#18, plotted in red).

\subsection{Proper motions}
\label{sec:pm}

Because proper motions are determined with AF, the impact of crowding is less
severe than on BP/RP spectro-photometry, or on RVS spectra. Figure~\ref{fig:hpm}
shows the histogram of the differences between the final proper motion --
including all error sources -- and the initial error-free simulated one, for the
known cluster members. The median systemic proper motion can be recovered with a
bias of 5--10~$\mu$as~yr$^{-1}$, i.e., with a systematic error of roughly
0.1--0.2\%. The intrinsic spread of the simulated GCs is of the order of
100~$\mu$as~yr$^{-1}$, depending on the GC. The observed spread is around
400--500~$\mu$as~yr$^{-1}$, if one includes all the simulated stars down to the
magnitude limit. It is almost entirely explained with the nominal post-launch
performances applied on top of the intrinsic spread, especially when including
faint stars. In fact (see Section~\ref{sec:crowd2}) the crowding errors {\em on
the affected stars} are in the worst cases a few 100~$\mu$as~yr$^{-1}$ for the
centroiding determination, and $\simeq$30~$\mu$as~yr$^{-1}$ for chromaticity
errors caused by G$_{\rm{BP/RP}}$ crowding errors. 

\begin{figure}
 \centering
 \includegraphics[width=0.9\columnwidth]{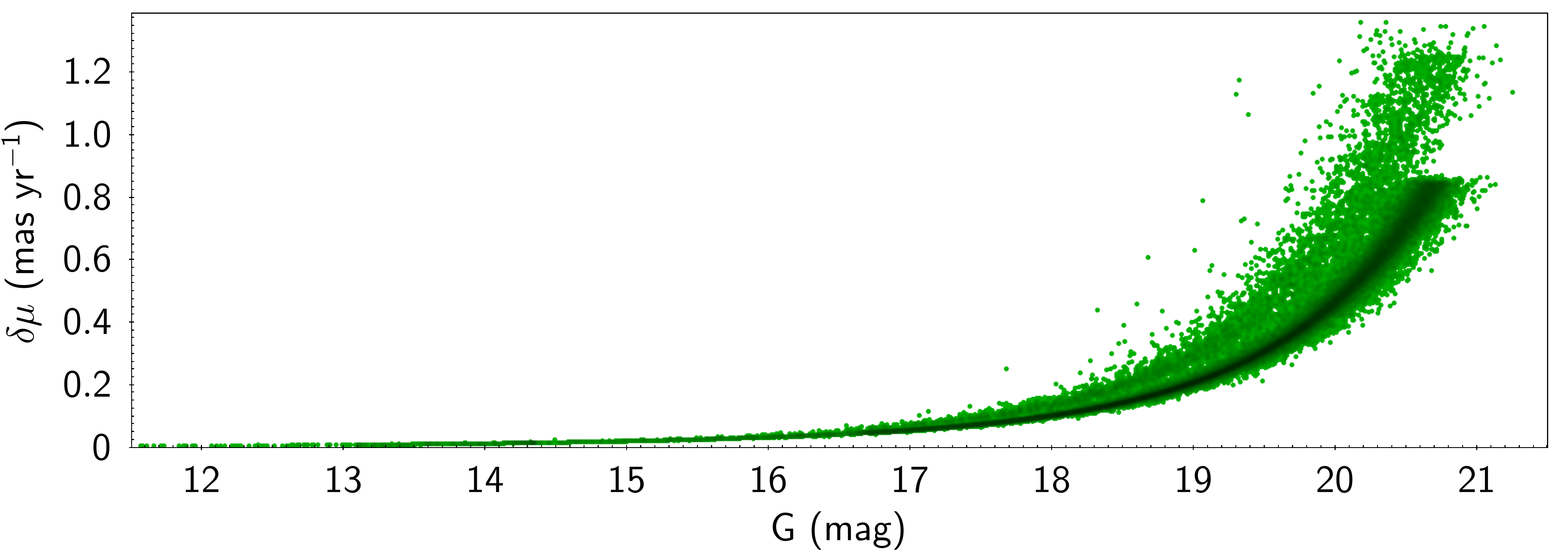}
 \includegraphics[width=0.9\columnwidth]{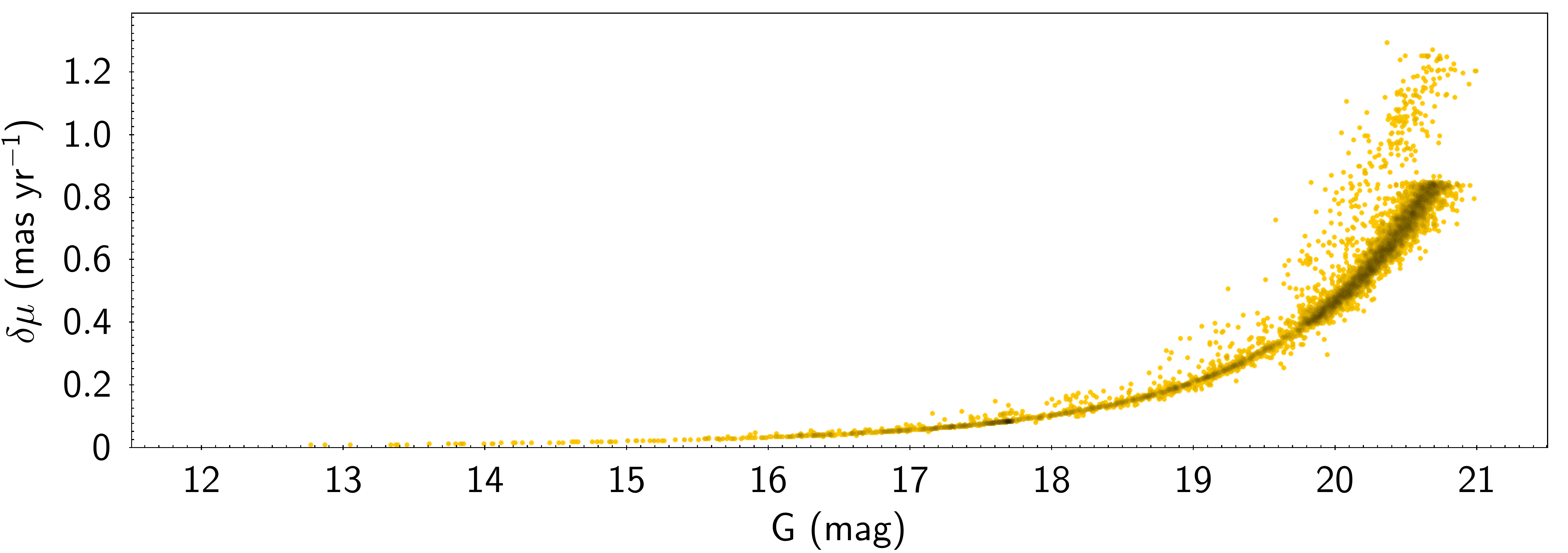}
 \includegraphics[width=0.9\columnwidth]{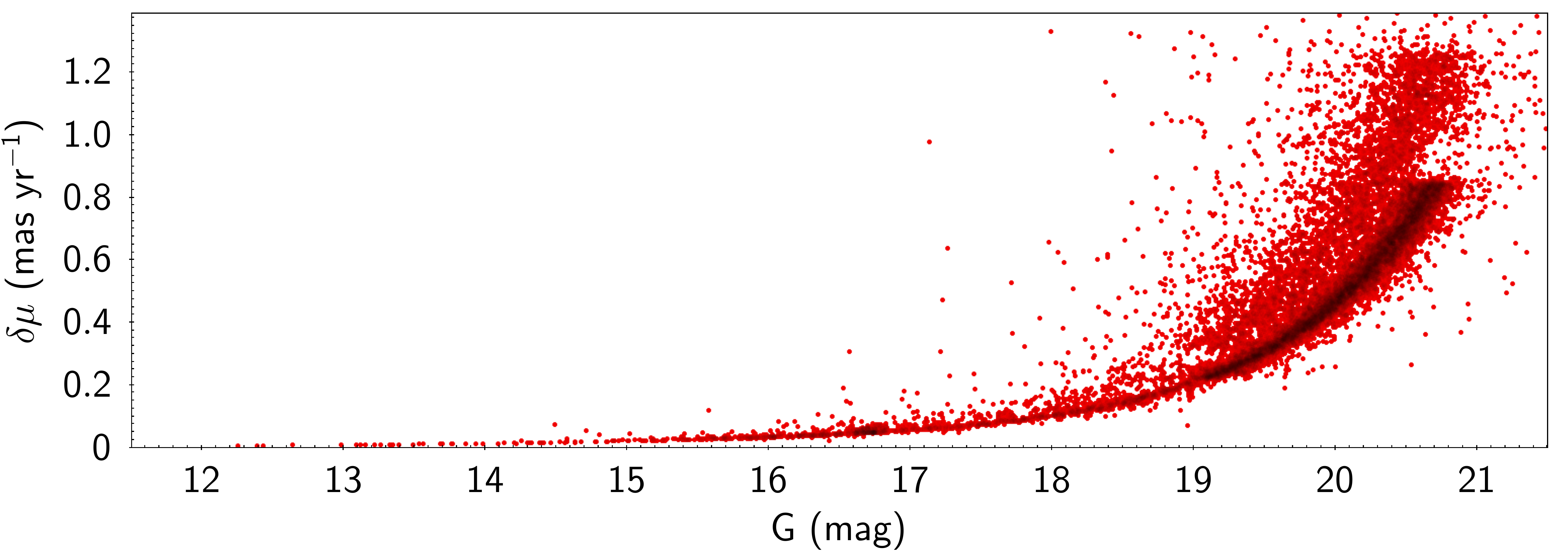}
 \caption{Final simulated proper motion errors, including the effect of crowding)
 for the easy (green points), intermediate (yellow points), and difficult cases
 (red points), as a function of final simulated G magnitude. Only cluster members
 are shown.}
  \label{fig:epm}
\end{figure}

\begin{figure*}
 \centering
 \includegraphics[width=0.33\textwidth]{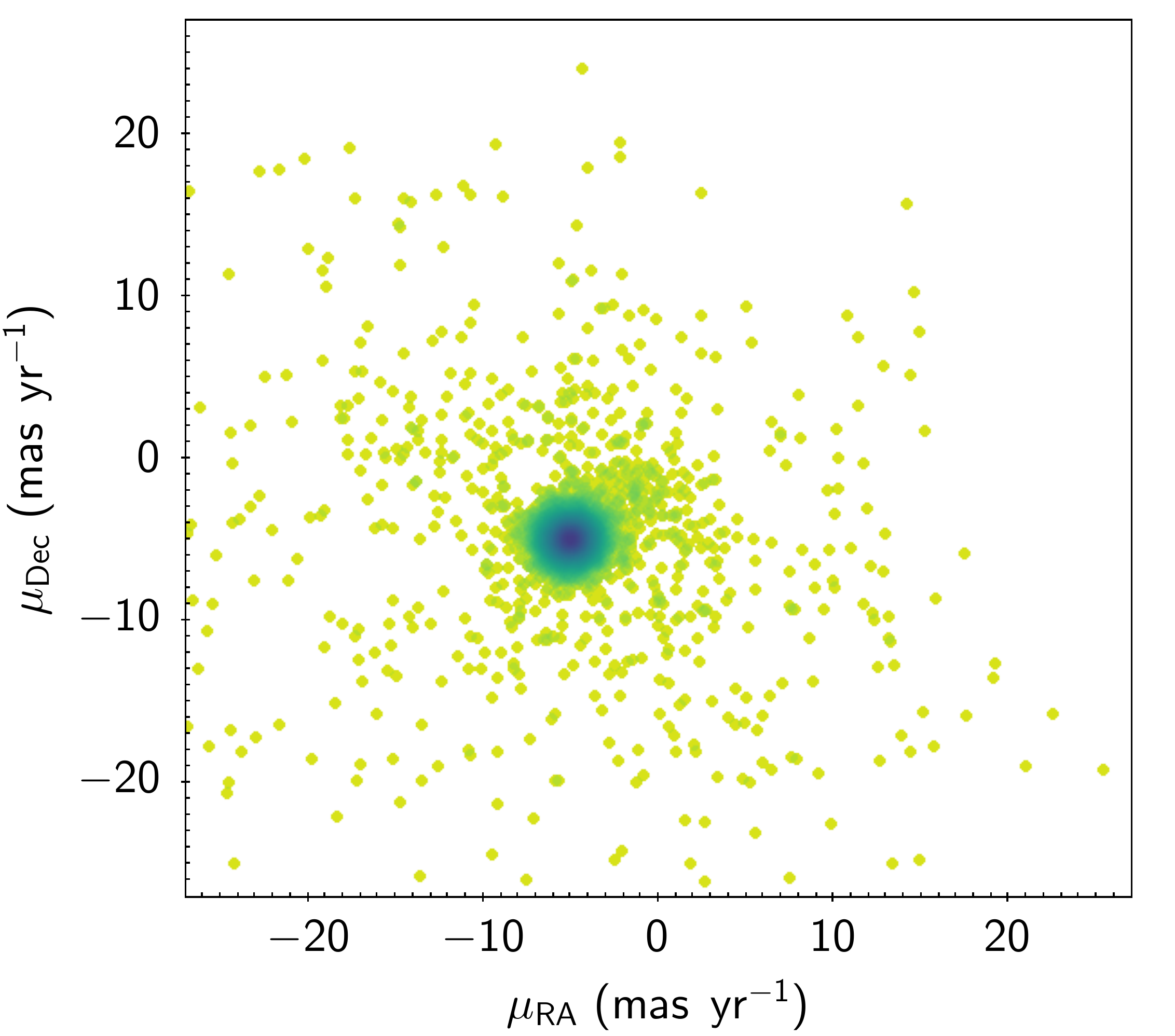}
 \includegraphics[width=0.33\textwidth]{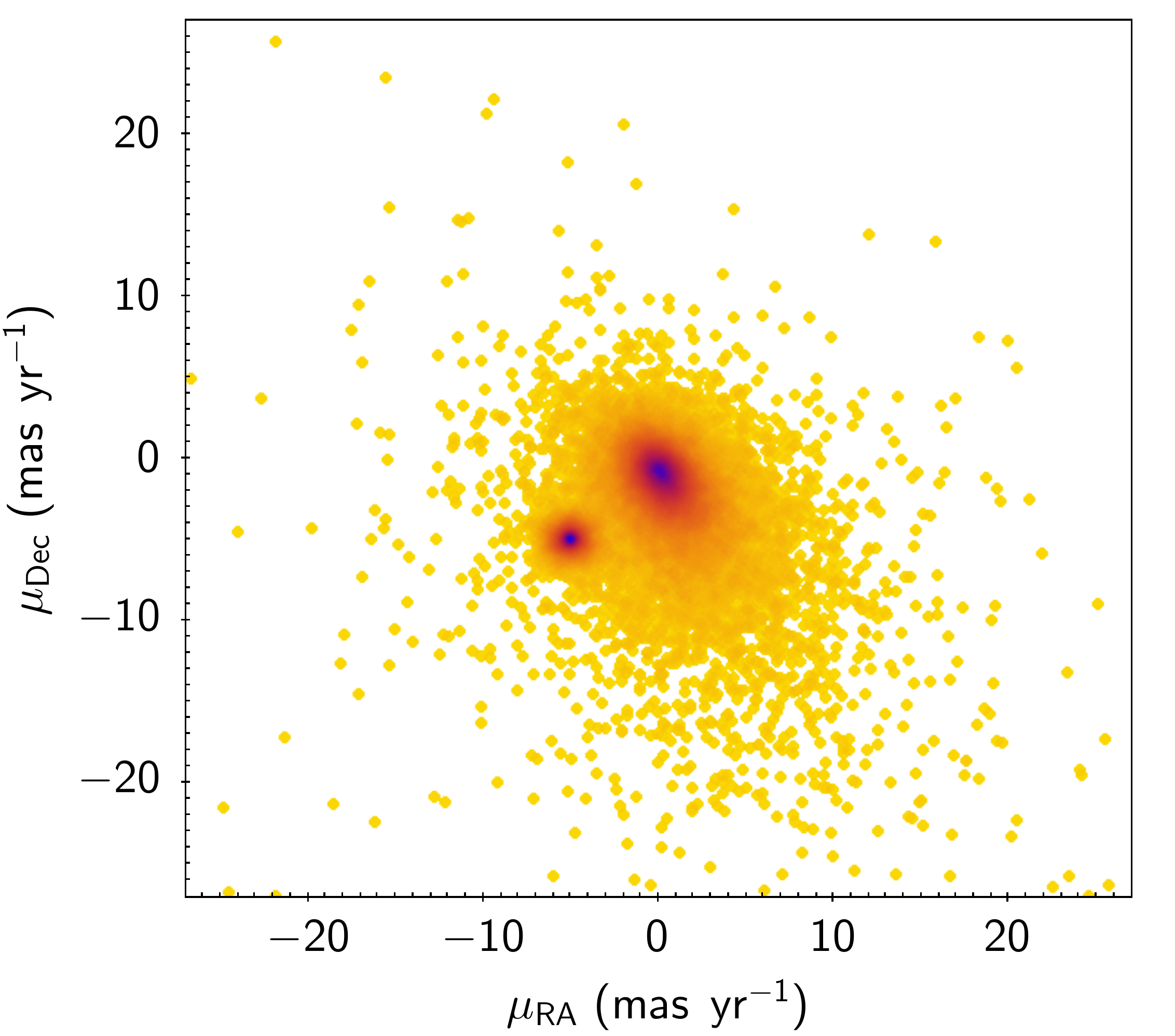}
 \includegraphics[width=0.33\textwidth]{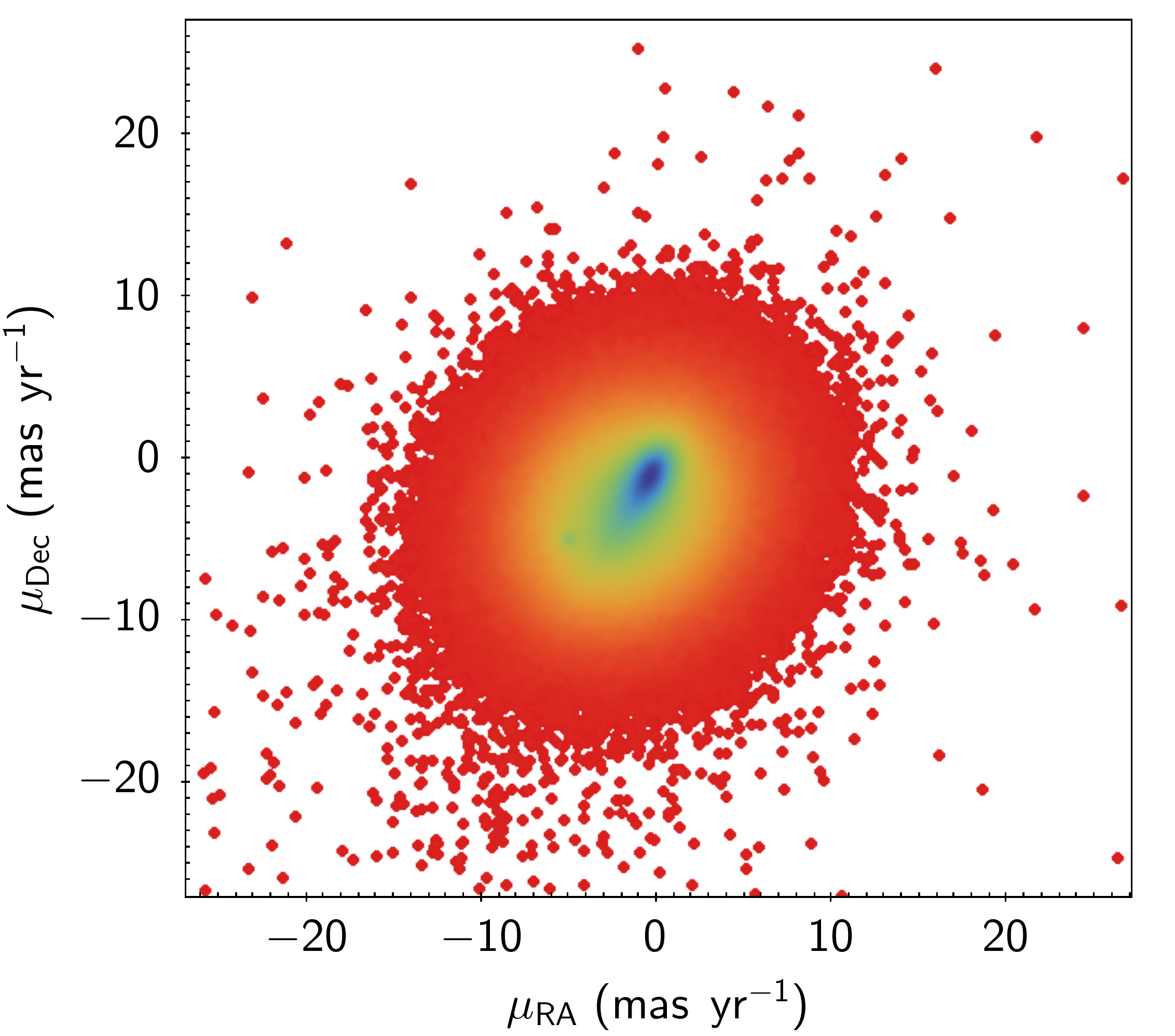}
 \caption{Proper motion diagrams for all the stars, cluster and background
 members, in the easy (green points), intermediate (yellow points), and
 difficult cases (red points). The colour scale refers to the density of points,
 with darker toner referring to higher densities. The cluster lies at (--5,--5)
 in all diagrams, and it is always easily recognized, but it is overwhelmed by
 field stars in the last case.}
  \label{fig:pm}
\end{figure*}

Concerning individual stars, Figure~\ref{fig:epm} shows the behaviour of the
final simulated errors as a function of G magnitude. It can be useful to compare
these simulations with the first results from the HST PROMO project for M15
\citep[Hubble Space Telescope PROper MOtion project,][]{bellini14}. Their typical
error around the G=15~mag is 5--10 times higher than the error expected from
Gaia, about 0.1~mas~yr$^{-1}$, but at G=21~mag they are about half the Gaia
error, at 0.5~mas~yr$^{-1}$. Moreover, HST proper motions can reach several
magnitudes below the Gaia limit. On the other hand, the advantage of Gaia's
proper motions is that they cover a larger area around each GC -- actually, the
whole sky -- and that they are {\em absolute} proper motions. 

We show in Figure~\ref{fig:pm} the typical vector diagrams of all simulated
stars, including faint stars and background contaminants. The clusters are
always easily separated from the field population, except in the case of the
extreme bulge background. In the following sections, we will select the probable
members with a {\em loose} criterium, i.e., stars within 1~mas~yr$^{-1}$ from
the systemic GC motion, and with a {\em strict} criterium, i.e., within
0.3~mas~yr$^{-1}$ from the systemic GC motion. 

\subsubsection{Measurement of derived quantities}

\begin{figure}
 \centering
 \includegraphics[width=0.9\columnwidth]{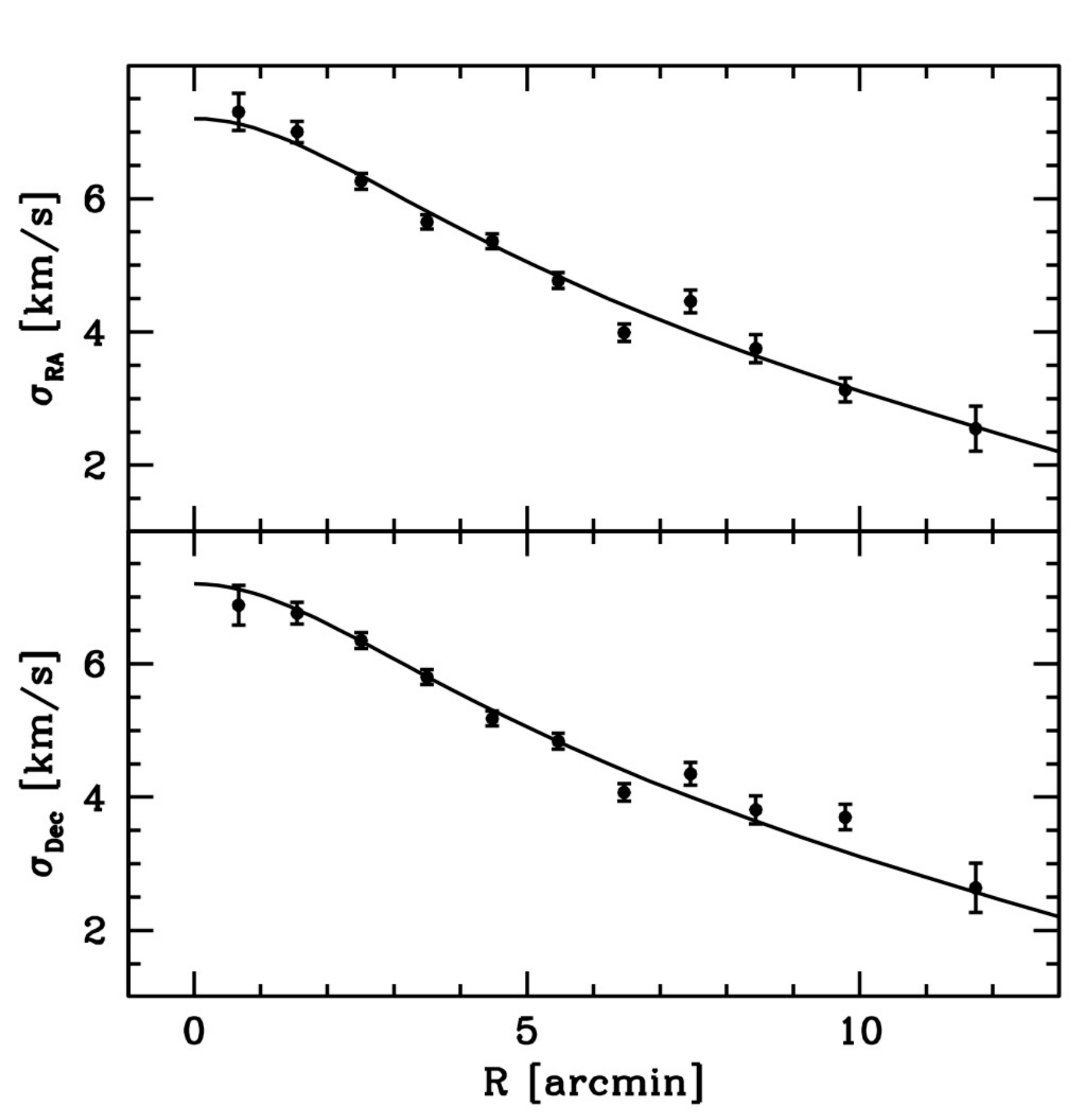}
 \caption{RV dispersion profile in RA and Dec for the easy case cluster.}
  \label{fig:sig}
\end{figure}

To assess the derived quantities that will be measurable from Gaia data, we
performed a basic test on the easy case GC. We excluded all stars that were
clearly unrelated to the GC with a very loose proper motion selection
($\pm$3~mas~yr$^{-1}$) and all the classic blends. We then converted proper
motions in the RA and Dec directions into velocities, using the GC distance and
propagating its error (see Section~\ref{sec:par}). We then estimated the RV
spreads $\sigma_{\rm{RA}}$ and $\sigma_{\rm{Dec}}$ as a function of distance
from the GC center with their errors. We used the maximum likelihood estimation
method (MLE) with the likelihood formulation described by \citet{pryor93},
\citet{walker06} and \citet{martin07}, which takes into account the errors on
measurements as well. The actual errors on the RV dispersions in each
radial bin are of about 0.4--1.4~km~s$^{-1}$. The result is displayed in
Figure~\ref{fig:sig}, showing that Gaia will be able to determine the radial
profile of the RV dispersion of nearby GCs with errors of about 1~km~s$^{-1}$ in
the nearest GCs.

\begin{figure*}
 \centering
 \includegraphics[width=0.33\textwidth]{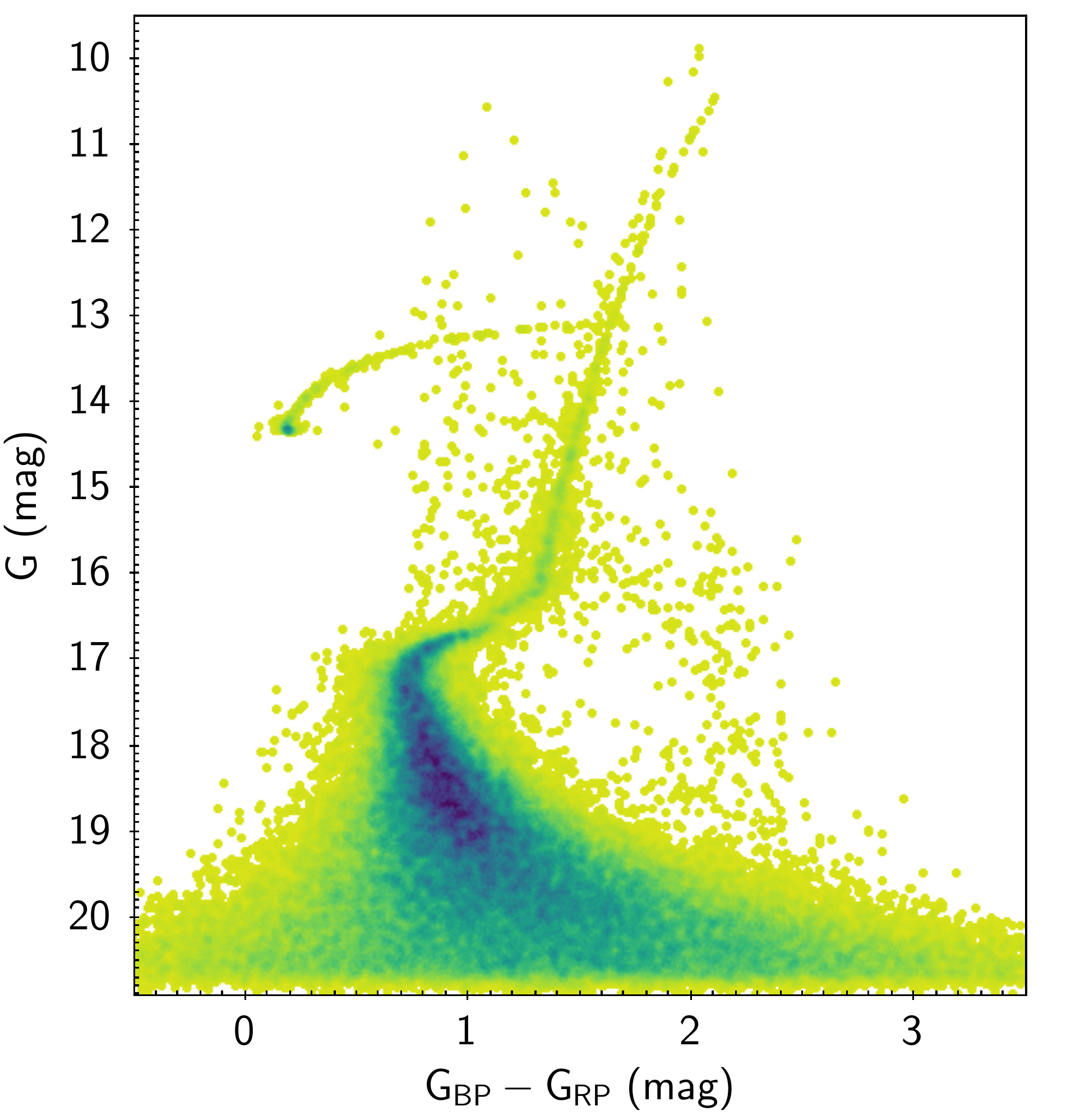}
 \includegraphics[width=0.33\textwidth]{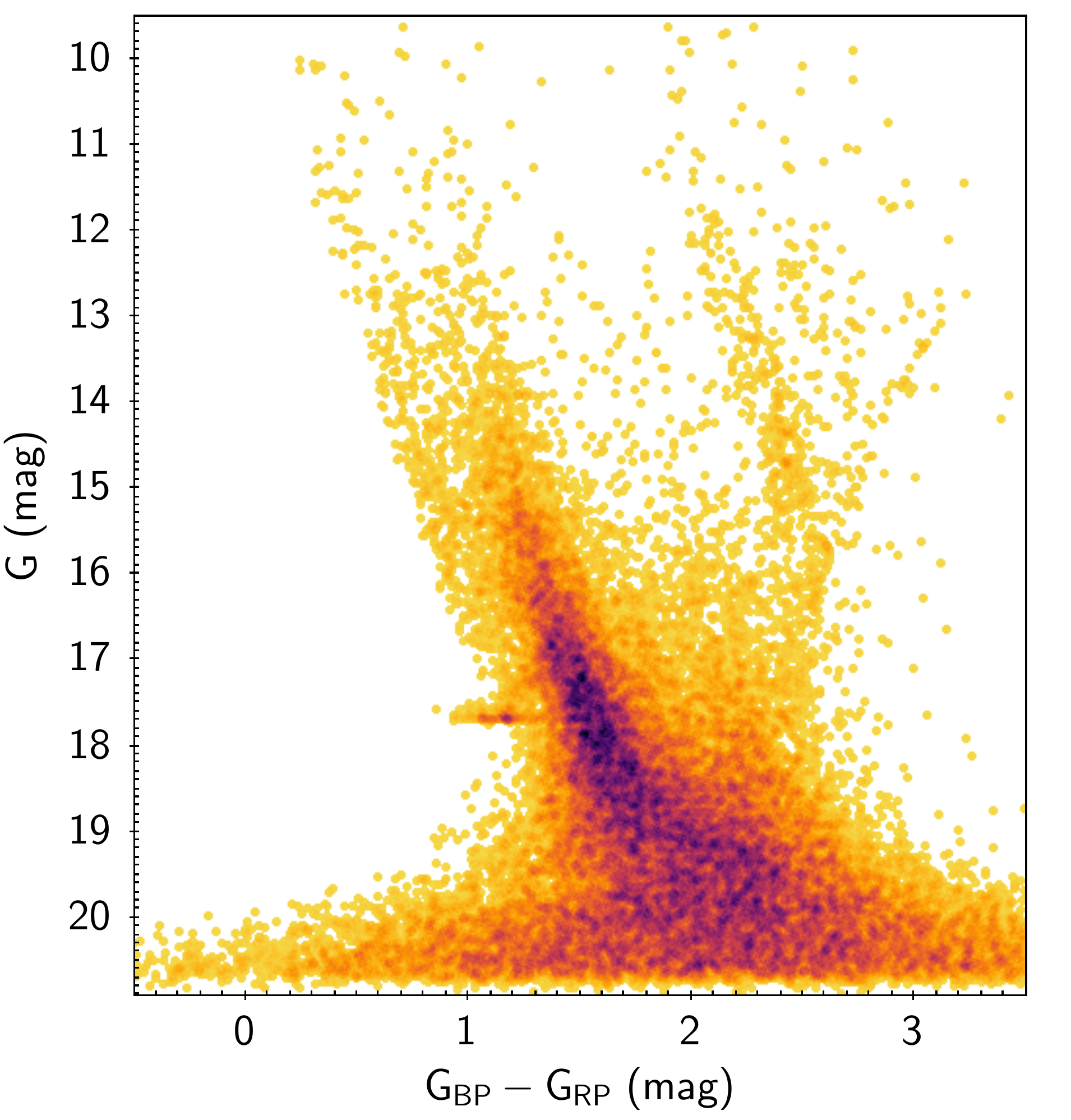}
 \includegraphics[width=0.33\textwidth]{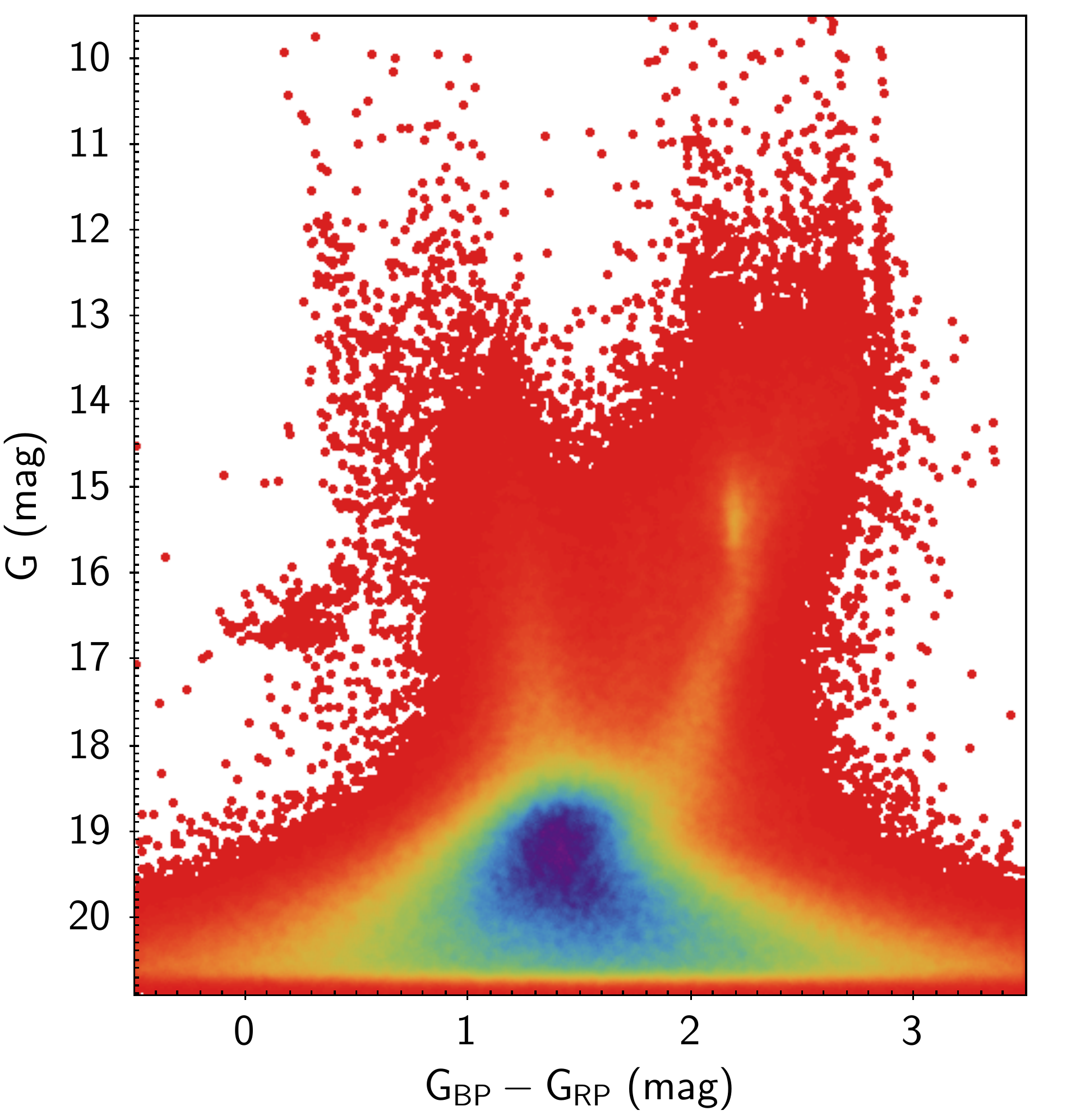}
 \includegraphics[width=0.33\textwidth]{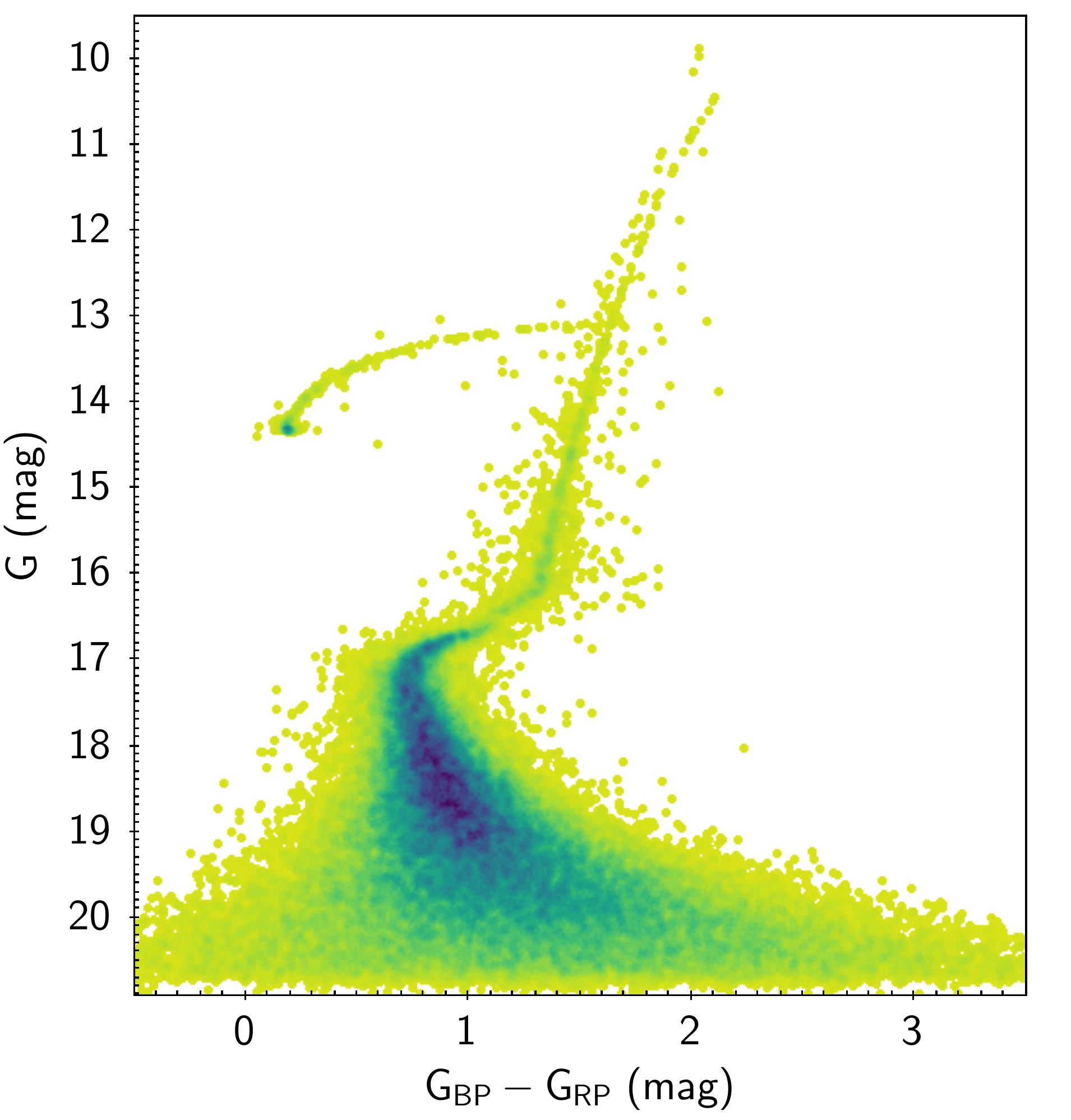}
 \includegraphics[width=0.33\textwidth]{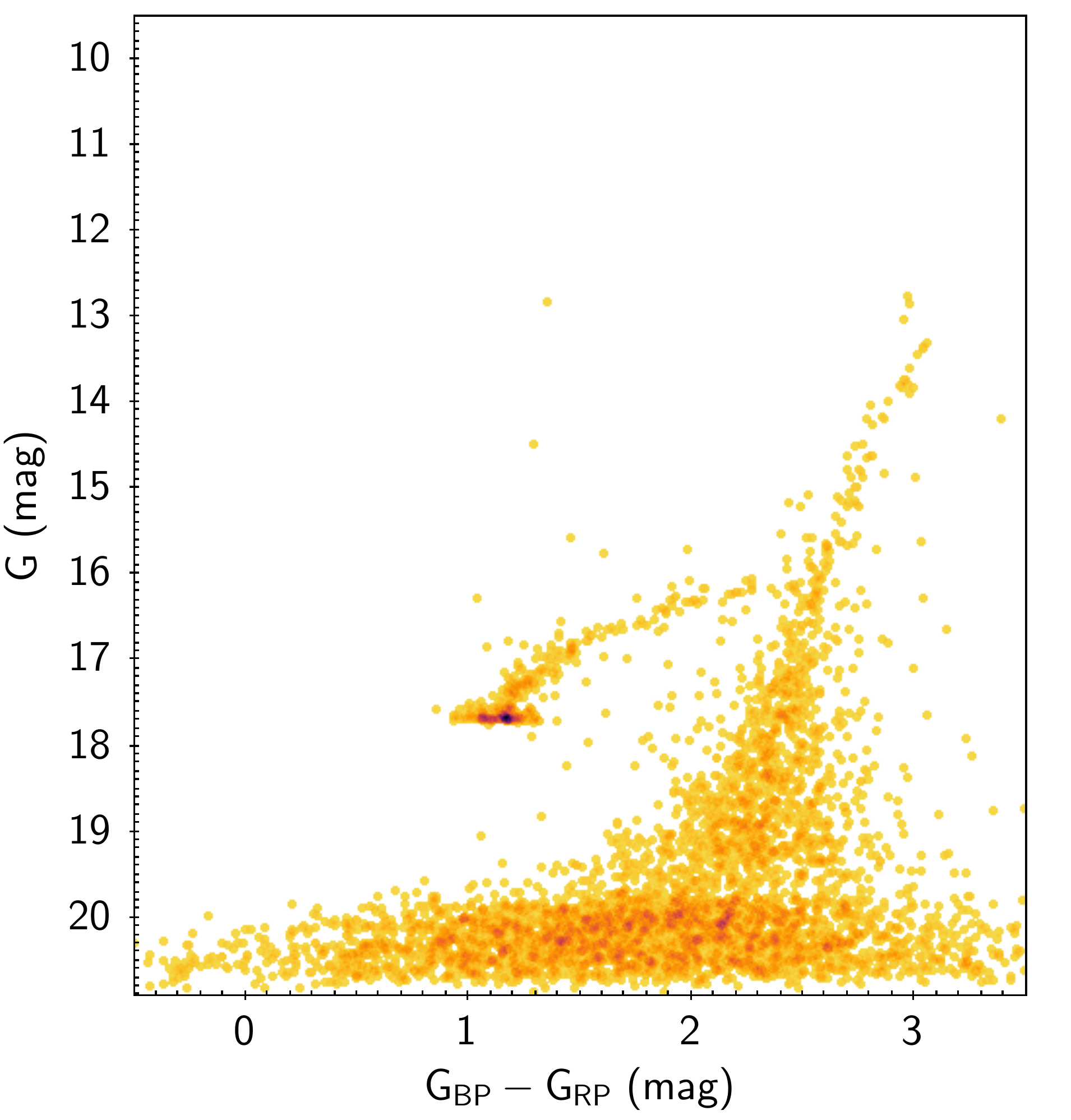}
 \includegraphics[width=0.33\textwidth]{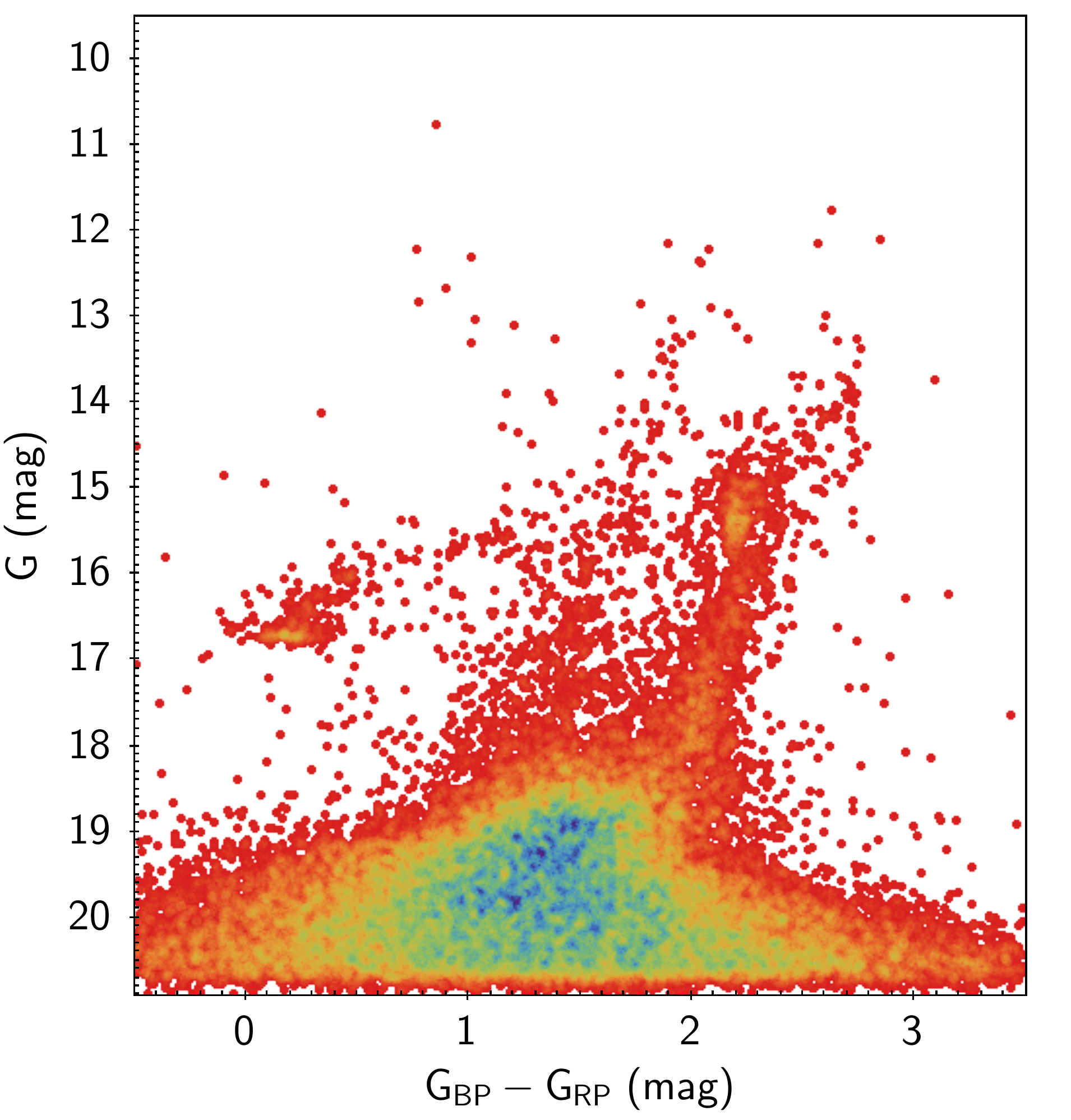}
 \caption{The simulated Gaia photometry for the three example GCs. Top panels
 show all the simulated stars, regardless of membership. Bottom panels show only
 the loosely selected probable members (see text for more details). The left
 panels refer to the easy case, the center ones to the intermediate case, and the
 right ones to the difficult case. The colour scale refers to the density of points,
 with darker toner referring to higher densities. }
  \label{fig:cmd}
\end{figure*}

More in general, we can say that:

\begin{enumerate}
\item{Proper motions obtained by Gaia in GCs will allow the study of dynamical
relaxation in the 5--10 closest GCs (see also Section~\ref{sec:phot} and
Figure~\ref{fig:cmd}). This is because at least a few magnitudes below the GC
turn-off point are required to adequately sample a range of stellar masses.
Figure~\ref{fig:cmd} clearly shows that this is possible only for clusters at
$\simeq$5~kpc or less, because of the Gaia detection limit. Therefore,
these studies will have to rely on deeper proper motion data sets for more
distant GCs\footnote{Deeper measurements could be obtained if the mission
lifetime is extended beyond its nominal 5-year duration. The SEA pipelines also
have the potential of recovering faint stars. Finally, the use of the 2D images
obtained for some crowded fields will help by directly recovering some faint
stars and by improving the deblending pipelines, thus recovering more faint
stars also in other regions.}.}
\item{Two past studies dealt with the determination of masses in GCs with Gaia
data \citep{an12,sollima15} yielding somewhat conflicting results. The main
problem is to break the mass-anisotropy degeneracy, with the anisotropy signal
being stronger outside the central regions where relaxation has the highest
effect. Therefore, blends by cluster members are irrelevant there, as discussed
in Section~\ref{sec:crowdeval} and shown in Figure~\ref{fig:counts}. And even in
the case of heavy background contamination, proper motions will be only mildly
affected (Figure~\ref{fig:epm}). The assumption by \citet{sollima15} that one
can only use stars contaminated by less than 10\% of the flux by objects more
distant than 3.54", is far too pessimistic. The simulations by \citet{an12},
prove that 10\% mass estimates can be provided for GCs if one can count on at
least 100 tracers (we can use red giants) with proper motion errors
$<$100~$\mu$as. As shown in Table~\ref{tab:clusters} and Figure~\ref{fig:epm},
the number of stars per GC with these errors will be more like a few
10$^3$--10$^4$, except maybe for the most extreme cases of bulge clusters.
Moreover, our test with the easy GC above shows that RV spreads can be measured
as a function of distance from the center with 1~km~s$^{-1}$ errors at 5~kpc.
Therefore, as concluded by \cite{an12}, it will be possible to determine GC
masses with an error of 10\% or less at least up to 15~kpc. This requires
ground-based RVs to complement Gaia proper motions\footnote{For the closest GCs,
Gaia will provide some good RVs as well, see also Section~\ref{sec:rvsres}, for more
details.}.}
\item{Gaia will be able to provide the necessary quantities for computing
accurate GC orbits within the Galaxy, because, as we saw, it will provide
high-quality systemic proper motions for each GC, with a bias of
a few $\mu$as~yr$^{-1}$ and errors of the order of 1\%. Distances will be
obtained with similar errors, as discussed below (Section~\ref{sec:par}).} 
\item{Once the orbits are obtained they could in turn be used to look for extra
tidal stars and tidal tails. Even if the G$<$20.7~mag limit reduces the density
of targets that will be actually measured by Gaia around GCs, the proper motion
signal of escaped GC stars should be very sharp, and in most cases significantly
different -- within the uncertainties -- from that of the surrounding field
population (see also Figure~\ref{fig:pm}).}
\item{Gaia could be able to help in the open and hotly debated problem of
multiple populations \citep{kraft94,gratton04}, in particular by revealing
kinematical differences among different GC sub-populations, if they are present.
For example, \citet{bellazzini12} reported the possible detection of a different
rotation pattern between the Na-rich and Na-poor stars in some of the GCs in
their sample, with rotation amplitude differences not larger than
$\simeq$5~km~s$^{-1}$. Depending on the distance and the inclination of the
rotation axis, this would correspond, in terms of proper motions, to a
difference in rotatation of the order of 1~mas~yr$^{-1}$, which would be
detectable for those GCs having at least 100 stars with measurement errors of a
few 10~$\mu$as.}
\end{enumerate}

\begin{figure}
 \centering
 \includegraphics[width=\columnwidth]{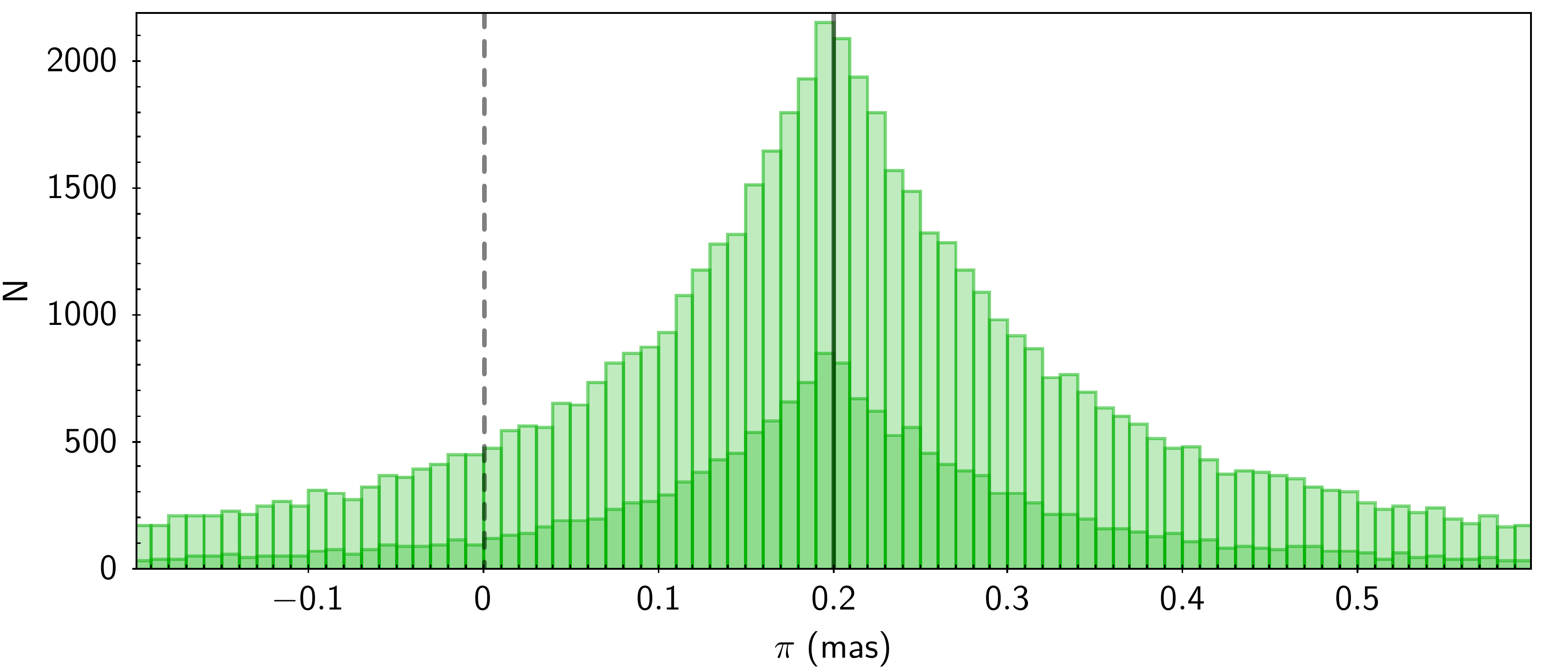}
 \includegraphics[width=\columnwidth]{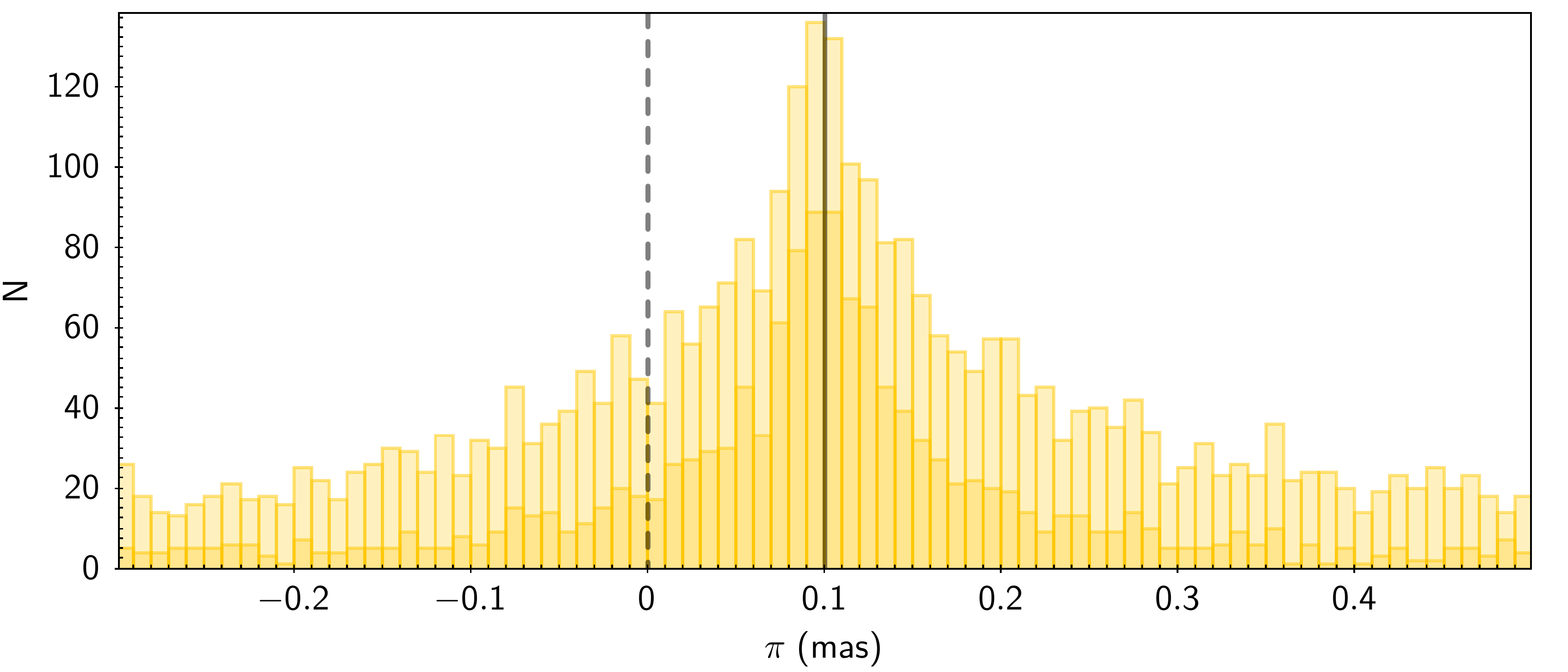}
 \includegraphics[width=\columnwidth]{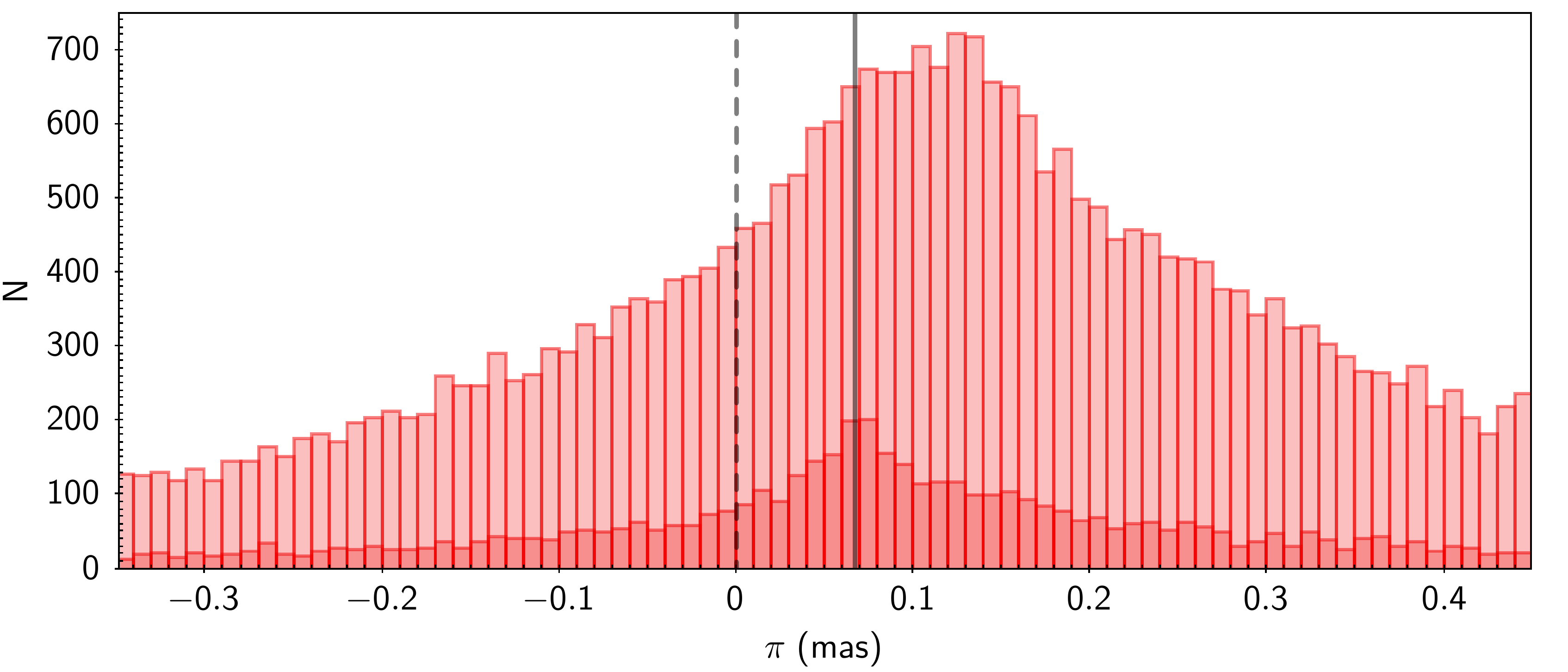}
 \caption{Histograms of the parallaxes for the easy (top panel), intermediate
 (middle panel), and difficult (lower panel) clusters. The vertical dotted line
 in all panels marks the zero parallax limit, while the solid line marks the true
 parallax of each simulated GC. Lighlty shaded histograms show the member stars
 selected with the loose criterium and heavily shaded ones with the strict
 criterium (see Section~\ref{sec:pm}).}
  \label{fig:par}
\end{figure}

\begin{table}
\caption{Systemic parallax for the simulated clusters, based 
on members selected with the strict and loose criteria (see
text for more details). The columns contain: {\em (1)} the cluster
number; {\em (2)} concentration parameter; {\em (3)} projected 
cluster distance and {\em (4)} parallax; {\em (5)} background 
type; {\em (6)} systemic parallax and error for members selected 
with the strict criterium and {\em (7)} with the loose criterium.
 \label{tab:dist}}
\begin{center}
\begin{tabular}{lcrrrrrrrrllll}
\hline
\\
Cluster & c & D$_0$ & $\varpi_0$ & field & 
$\varpi_{\rm{strict}}$ & 
$\varpi_{\rm{loose}}$ & \\
        &   & (kpc) & ($\mu$as) &   &     ($\mu$as) &  ($\mu$as) \\
\hline
\\
\# 1  & 1.0 &  5 & 200.0 & halo  & 200.2$\pm$2.6 & 200.2$\pm$0.6 \\
\# 2  & 1.0 & 10 & 100.0 & halo  & 101.6$\pm$2.1 &  99.3$\pm$0.8 \\
\# 3  & 1.0 & 15 &  66.7 & halo  &  69.2$\pm$2.6 &  69.1$\pm$1.0 \\
\# 4  & 1.0 &  5 & 200.0 & disk  & 203.9$\pm$6.1 & 197.4$\pm$1.3 \\
\# 5  & 1.0 & 10 & 100.0 & disk  &  95.6$\pm$4.4 &  98.5$\pm$1.5 \\
\# 6  & 1.0 & 15 &  66.7 & disk  &  74.0$\pm$5.6 &  65.8$\pm$1.9 \\
\# 7  & 1.0 &  5 & 200.0 & bulge & 206.8$\pm$2.3 & 200.8$\pm$0.6 \\
\# 8  & 1.0 & 10 & 100.0 & bulge & 100.1$\pm$1.9 &  99.1$\pm$0.6 \\
\# 9  & 1.0 & 15 &  66.7 & bulge &  63.9$\pm$2.3 &  67.9$\pm$0.7 \\
\# 10 & 2.5 &  5 & 200.0 & halo  & 197.5$\pm$0.7 & 198.4$\pm$0.4 \\
\# 11 & 2.5 & 10 & 100.0 & halo  & 101.4$\pm$1.3 & 101.2$\pm$0.6 \\
\# 12 & 2.5 & 15 &  66.7 & halo  &  62.2$\pm$1.5 &  66.4$\pm$0.9 \\
\# 13 & 2.5 &  5 & 200.0 & disk  & 200.0$\pm$1.6 & 200.2$\pm$0.8 \\
\# 14 & 2.5 & 10 & 100.0 & disk  & 100.0$\pm$3.1 & 102.8$\pm$1.3 \\
\# 15 & 2.5 & 15 &  66.7 & disk  &  74.3$\pm$3.2 &  70.5$\pm$1.8 \\
\# 16 & 2.5 &  5 & 200.0 & bulge & 199.3$\pm$0.7 & 198.6$\pm$0.4 \\
\# 17 & 2.5 & 10 & 100.0 & bulge & 101.6$\pm$1.3 &  99.5$\pm$0.5 \\
\# 18 & 2.5 & 15 &  66.7 & bulge &  65.8$\pm$1.4 &  67.0$\pm$0.7 \\
\hline
\end{tabular}
\end{center}
\end{table}

\subsection{Parallaxes}
\label{sec:par}

The major problem in deriving distances from parallaxes, especially when the
error is comparable to the measurement, is that the errors on distances are not
symmetric: by simply inverting the parallax one could obtain a very wrong
distance estimate \citep{bailer15}. Statistically speaking, it is also possible
to obtain negative parallaxes, as illustrated in Figure~\ref{fig:par}, which
would result in meaningless distance estimates if not treated properly.

When one is interested in the systemic distance of a GC, however, negative
parallaxes are useful to assess the uncertainty on the measurements and their
distribution. Figure~\ref{fig:par} shows the distribution of parallaxes for
probable members, as defined in Section~\ref{sec:pm}. As can be seen, not only
the average parallax of the clusters is very well recovered, but also its
uncertainty, provided that one makes use also of the negative parallaxes. The
exception is the case of the difficult cluster, which has a background
contamination that overwhelms the signal from the cluster members. As can be
seen, the recovered parallax is strongly biased by the background contaminants.
A more strict members selection apparently produces better results (but see
below). However, the resulting distribution of strict members is still
asymmetric, because of the blends between cluster and field stars. 

This type of problem is not specific of Gaia data, of course, but it is
extremely important to employ a reliable membership selection and robust
statistical modeling of the parallax distribution, which can be skewed and is
heteroscedastic. We employed an MLE analysis of the 18 simulated GCs using the
strict and loose membership criteria described above. We used the likelihood
estimator described by \citet{pryor93}, \citet{walker06}, and \citet{martin07},
which takes into account the highly variable errors, and we computed the most
probable parallax $\varpi$ and its error $\delta\varpi$. The result is displayed
in Table~\ref{tab:dist}. We first note that the depth of a GC, as measured with
parallaxes, can correspond at most to a few~$\mu$as for the nearest GCs that
have large radii, even if one considers the tidal radius, and in the majority of
cases is well below 1~$\mu$as. Thus the observed spread is caused almost
entirely by measurement errors. The bias itself, i.e., the difference between
the true input parallax and the recovered one, is always 1\% or smaller, even
for the difficult case GC. The formal error on the recovered $\varpi$ is of the
same order. We also note that choosing a more restrictive membership selection
can often increase the formal errors without resulting in a better $\varpi$
determination. On the contrary, the bias is slightly increased when applying a
more restrictive membership selection \citep[see also][]{bailer15}. 

The scientific implications of $\simeq$1\% distances for most of the GCs in the
Galaxy are far reaching. They directly help in the determination and modeling of
GC orbits, that can then be used for dynamical studies of the MW, of the GC
themselves, and of the interactions between the two. For example they would help
simulations that use the GC orbits and tidal tails and streams to constrain the
Galactic potential \citep{penarrubia12,price14}, or they can help discerning the
origin of the GC themselves through dinamical modeling. But the area in which
significant breakthrough is expected lies in the determinations of stellar ages
\citep{gratton97} and masses \citep{feuillet16}. Distance has in fact a similar
effect as age when using high-quality CMDs (from HST or ground-based
observations) to estimate the {\em absolute ages} of GCs. By reducing the
distance determination errors by more than a factor of 10, we can expect a
significant improvement on the age determinations, obtaining absolute ages with
errors below 10\%. On the other hand, $\simeq$1\% distances can help in
constraining the stellar masses and more importantly, the surface gravities. When
trying to determine {\em relative ages} in GCs with multiple stellar populations,
one needs to disentangle the effects of chemical composition -- mainly Helium and
C+N+O abundance -- from age effects. A very accurate surface gravity
determination would remove one of the major sources of uncertainty in the
determination of stellar chemistry from spectroscopy.

\subsection{Photometry}
\label{sec:phot}

The Gaia colour magnitude diagrams (CMD) of the three example clusters are shown
in figure~\ref{fig:cmd}. The top panels show all the simulated stars, including
the background, while the bottom panels show only the probable members, selected
with the loose criterium described in the previous section. As can be seen,
background contamination is the first obvious cause of crowding errors for Gaia
photometry. A large reddening (like in the case of the disk cluster) or a
distance larger than 10~kpc also impact the quality of the CMD, because the
photometric errors, especially those on the G$_{\rm{BP}}$--G$_{\rm{RP}}$ colour,
increase quite rapidly with magnitude.

It will therefore not be possibile to reach the nominal photometric errors
promised by Gaia in those clusters that lie on a bulge or disk background.  This
is mainly caused by the extended shape of the BP and RP dispersed images, while
the G magnitudes will not suffer significantly from crowding. However, for
clusters that are relatively free from background contamination, the BP/RP
photometry of stars brighter than G$\simeq$15~mag will have an extremely good
quality, comparable to HST photometry. Fainter stars down to G$\simeq$18--19~mag
will still have a BP/RP quality that is comparable with the best ground-based
catalogues. 

While HST photometry will certainly be preferable for some applications, Gaia
photometry will have a few advantages: (1) the field of view of Gaia is not
limited by any field size: it covers the whole sky; (2) each star that has
Gaia astrometry from AF, also has BP/RP spectra from which rough stellar
parameters and a reddening estimate can be obtained \citep{gaia1}; (3) brighter
stars, generally red giants in the case of GCs, will also have RVs and more
accurate estimates of parameters and reddening from RVS
spectra\footnote{Accurate reddening estimates can also be obtained from diffuse
interstellar bands, that are included in the Gaia RVS wavelength range
\citep[see for example][and references therein]{puspitarini15}.}
\citep{kordopatis12}; and (4) the absolute photometric calibration of Gaia will
be based on one the largest, most homogeneous, and and most accurate set of
spectro-photometric standard stars to date
\citep{pancino12,altavilla15,marinoni16}, that will grant an accuracy of
$\simeq$1--3\% with respect to Vega \citep{bohlin04}. 

\subsection{Radial velocities}
\label{sec:rvsres}

\begin{figure}
 \centering
 \includegraphics[width=\columnwidth]{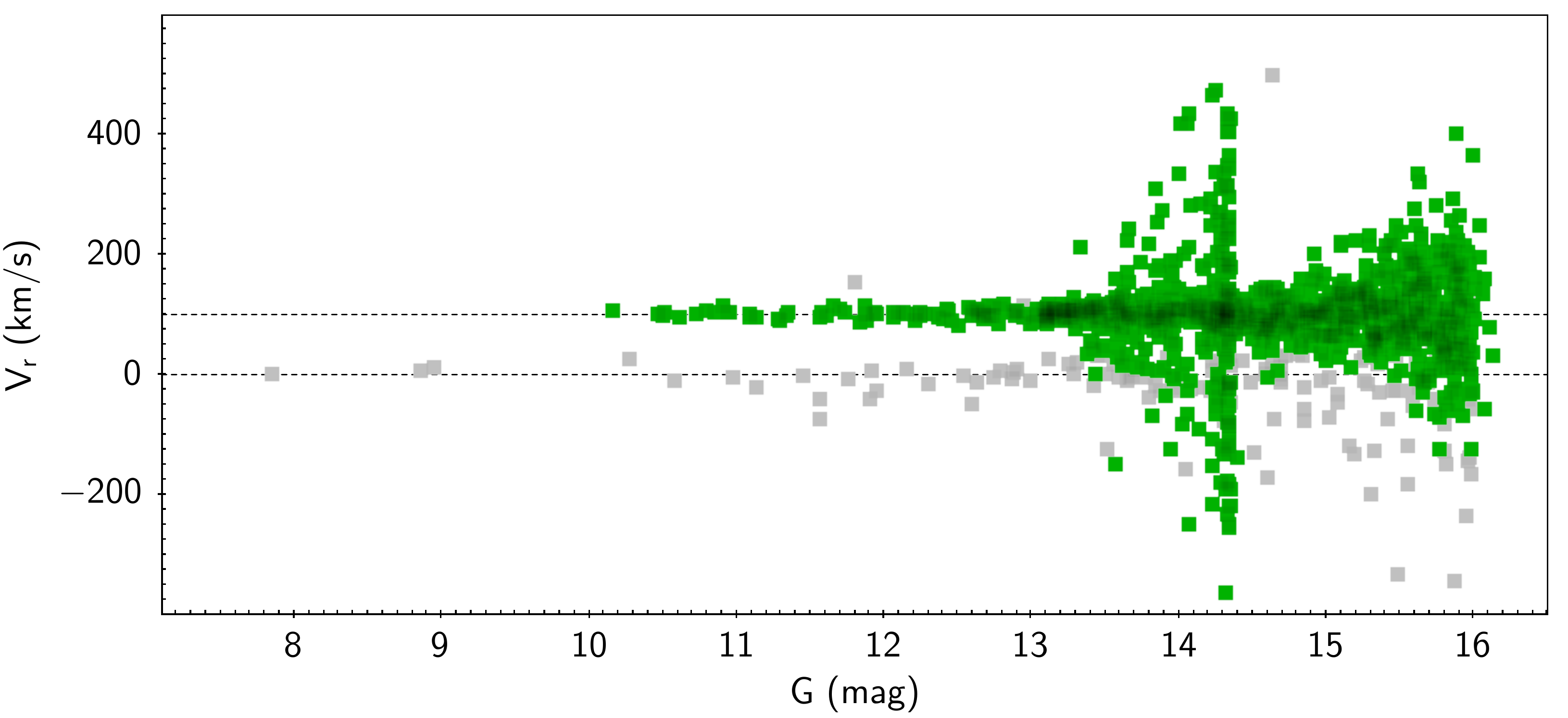}
 \includegraphics[width=\columnwidth]{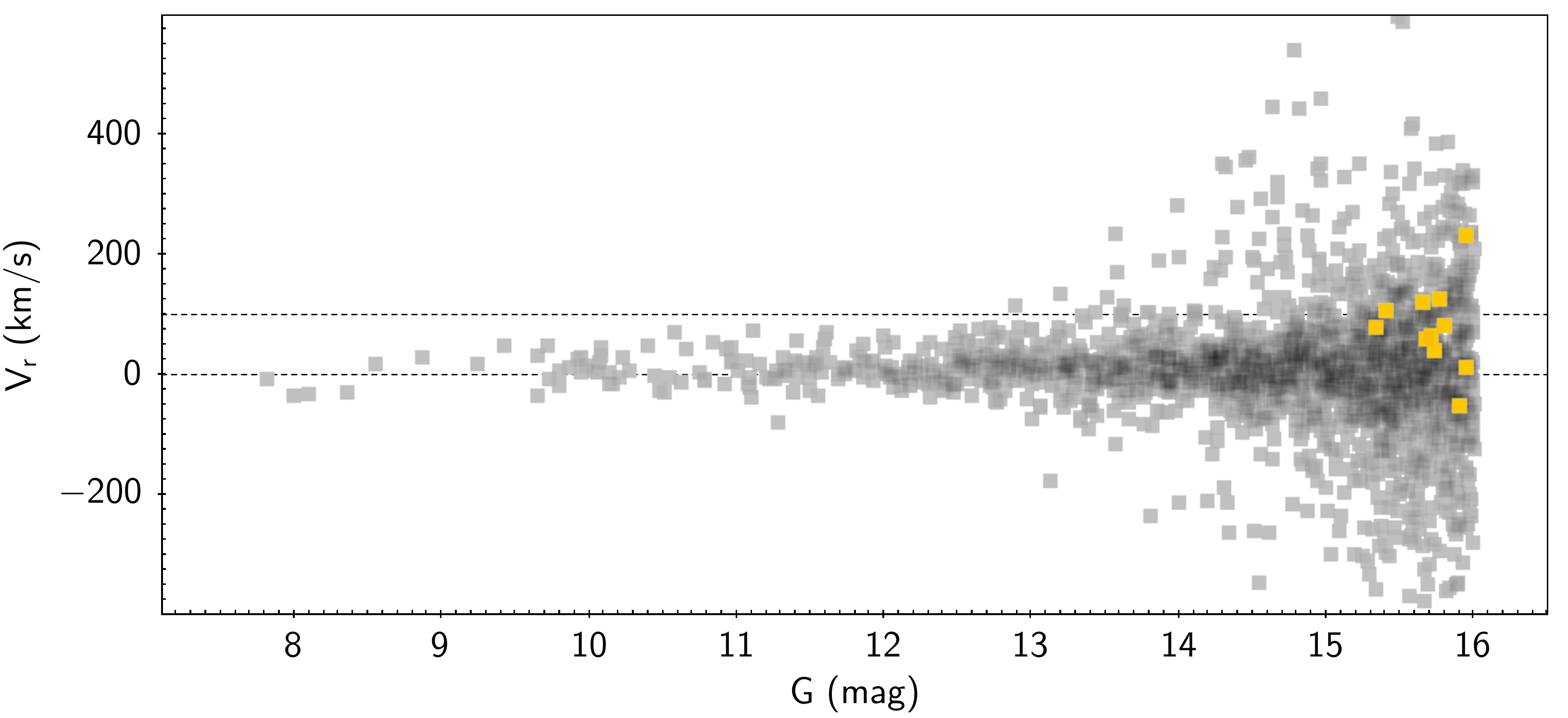}
 \includegraphics[width=\columnwidth]{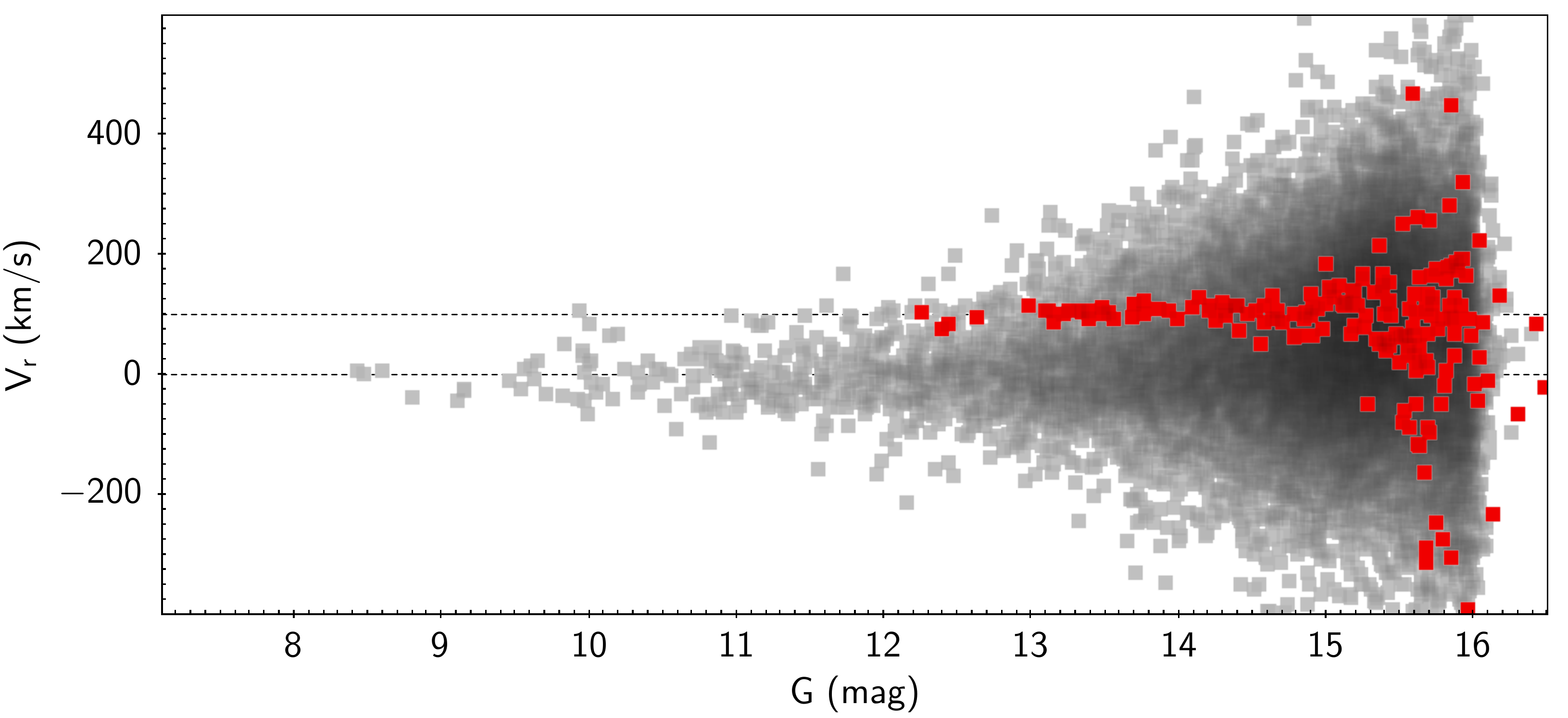}
 \caption{RV as a function of magnitude for the easy (top panel), medium (middle
 panel), and difficult (bottom panel) cases. The entire samples of available RV
 measurements are plotted in grey. The dotted lines mark zero and the true
 systemic velocity of the simulated clusters, 100~km~s$^{-1}$. Only true cluster
 members are coloured in  green, yellow, and red, respectively.} 
  \label{fig:vrad}
\end{figure}

Stars brighter than G$\simeq$17~mag will have RVs measurements. Gaia has already
produced billions of RVS spectra and by the end of the mission, each star will
be observed on average 40 times. This can be compared with the extremely
successful RAVE survey \citep[Radial VElocity Experiment,][]{rave}, which
measured RVs for half a million stars with V$<$12~mag. The Gaia end-of-mission
errors will vary with the star's colour and will be of the order of
1~km~s$^{-1}$ for the bright red stars (G$<$12.5~mag and cooler than F types)
and will be about 15--20~km~s$^{-1}$ or more for fainter and bluer stars.
Crowding will affect RVS more than any of the other instruments on board. Even
if the magnitude limit is brighter, the AL size of the spectra is more than an
arcminute on the sky (see Table~\ref{tab:numbers}). 

\begin{figure}
 \centering
 \includegraphics[width=\columnwidth]{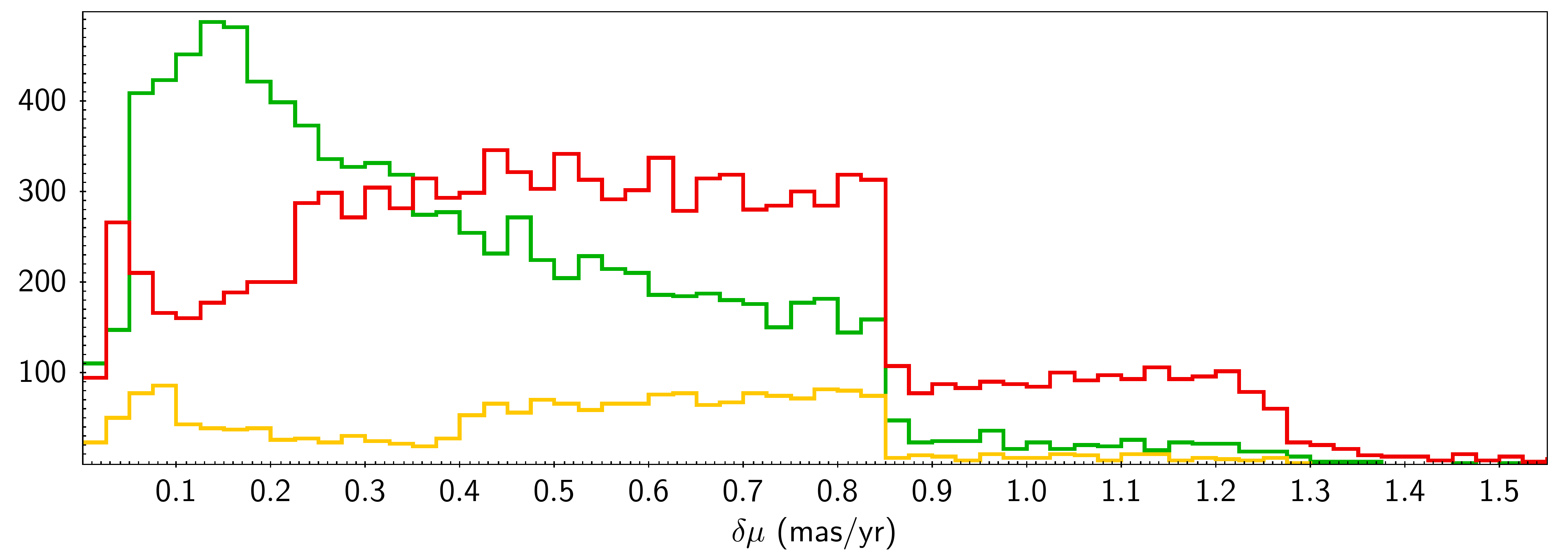}
 \includegraphics[width=\columnwidth]{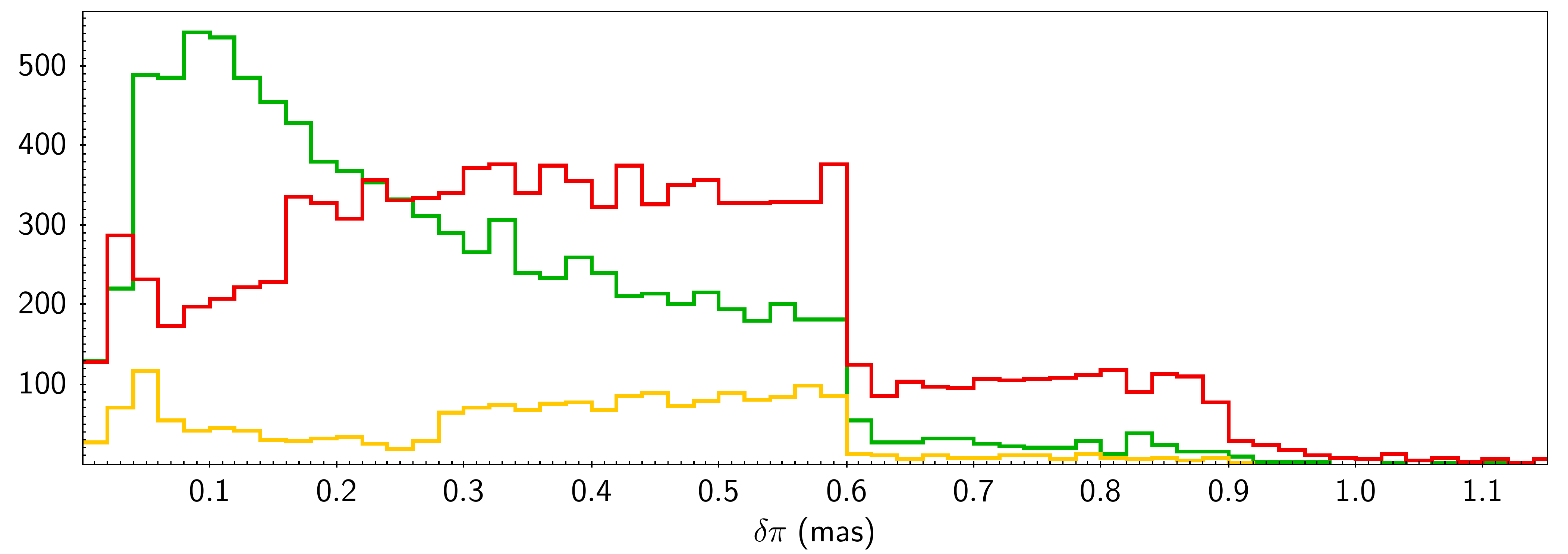}
 \includegraphics[width=\columnwidth]{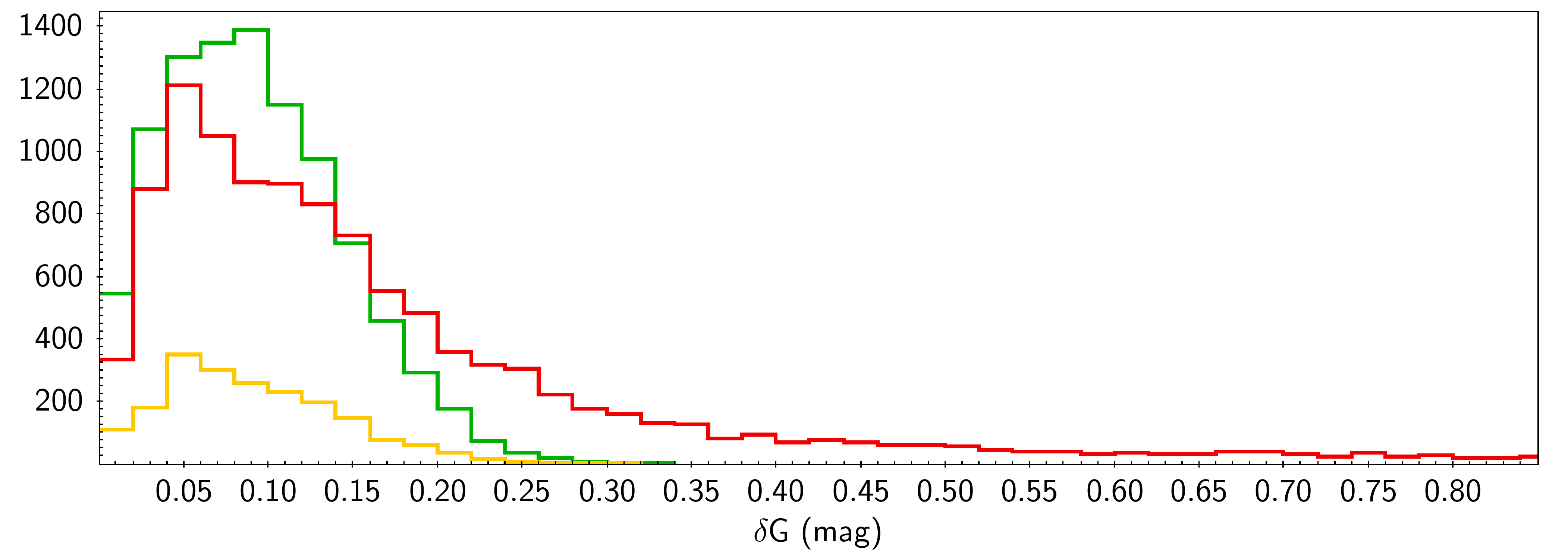}
 \caption{End-of-mission error distribution for all member stars (defined as in
 Figure~\ref{fig:vrad}) down to G=20.7~mag, within the central arcminute of the
 simulated GCs. The top panel shows proper motion errors, the middle one parallax
 errors, and the bottom ones G magnitude errors. The easy cluster is represented
 in green in all panels, the intermediate one in yellow, and the difficult one in
 red.} 
  \label{fig:core}
\end{figure}

Figure~\ref{fig:vrad} shows the simulated RV measurements, as a function of the
G magnitude, for the three example cases. Given the limitations of our
simulations, we immediately note that only a handful of reliable members in the
intermediate case have meaningful RV determinations, because of the combined
effect of high reddening -- for a GC -- and background contamination. We also
note the large scatter in RV for the faintest stars in the difficult case, where
background contamination takes its toll. However, for the easy case, we
notice that a sample of more than 100 stars with magnitudes above
$\simeq$13--14~mag is available, with errors around 1 or few km~s$^{-1}$. This
will happen for the closest 10--20 GCs. The stars with large errors between
13.5 and 14.5~mag are hotter horizontal branch stars. To conclude, we note that
a mission extension would increase the quality of the RVS spectra.

\subsection{The central arcminute}

As discussed in Section~\ref{sec:comp}, it is not feasible to simulate the degree
of completeness of Gaia data, because it varies across the sky, based on the
number of different passages and on their respective orientation. As discussed in
that section, nearby and relatively sparse GCs like $\omega$~Cen should be 100\%
complete down to V$\simeq$16~mag, even in the very core. This might appear
surprising, but it is a beneficial byproduct of the relatively shallow Gaia
magnitude limit. 

However, still the question remains: {\em how far into the GC core can we obtain
reliable measurements, at least for the stars that we will be able to measure?}
Figure~\ref{fig:core} shows the end-of-mission error distributions for all
member stars, down to G=20.7~mag, for the central arcminute of the simulated
clusters. As can be seen, astrometry is not too badly affected by crowding: the
maximum errors are of 1~mas~yr$^{-1}$ for proper motions and 0.7~mas for
parallaxes (to compare with Figure~\ref{fig:epm}). The majority of stars have
performances of the order of a few hundred $\mu$as~yr$^{-1}$ or $\mu$as, and
this including also the faintest stars. If this appears surprising, we recall
that what mostly governs the end-of-mission astrometric errors caused by
crowding in Gaia is the PSF size, which is comparable to that of HST.

The situation is different for BP/RP photometry or RVS, because even if the PSF
in the AC direction is still small, and thus allows to detect blends quite
efficiently, the AL size of the window is larger, and aligned with the dispersion
axis of the spectra. The performances in the central arcminute are in fact
significantly worse for these instruments than for the AF and the G band
magnitude in the case of crowded fields.

\section{Summary  and conclusions}
\label{sec:concl}

We used all the available information on the presently available Gaia deblending
pipelines, along with their results on simulated data, to model the behaviour of
Gaia in crowded areas. We computed additional crowding errors that were combined
with the post-launch science performances of Gaia and applied to a set of
simulated GCs with different concentration (c=1.0 and 2.5), distance (5, 10, and
15~kpc), and field population (halo, disk, and bulge).

We showed that:
\begin{enumerate}
\item{Gaia will provide estimates of the parallax and systemic motion of GCs with
unprecedented accuracies: $\simeq$1\% and $<<$1\%, respectively, for all GCs as
far as 15~kpc at least. Also, systemic RVs with errors of a few km~s$^{-1}$ will
be obtained for the 10--20 closest GCs, although our RVS simulations are not as
accurate as for the other Gaia instruments. This will allow for a very accurate
modeling of GC orbits.}
\item{The astrometry, obtained through the AF together with G magnitudes, will be
only marginally affected by crowding, having performances not too dissimilar from
the nominal Gaia performances even for the most field-contaminated bulge GCs.
This is an effect of the tiny Gaia PSF (0.17") in AF. The G magnitudes will have mmag performances and space quality to
G$\simeq$17--18~mag.}
\item{The proper motions of individual stars within GCs have sufficient quality
to obtain mass estimates with 10\% errors for GCs as far as 15~kpc at least, and
to identify small variations of the properties (rotations, spreads) of a few
km~s$^{-1}$ for GCs as far as 10~kpc.}
\item{The distances of GCs will be obtained with errors of $\simeq$1\%, and for
GCs heavily contaminated by field stars to a few percent. The impact on the
determinations of stellar masses and ages will be significant. It is expected
that GC absolute ages with errors below 10\% will be obtained. Also, better
estimates of surface gravities for GC stars with known distances will remove one
of the major uncertainty sources from abundance determinations with
high-resolution spectroscopy.}
\item{While it is difficult to simulate the exact completeness level of Gaia in
GCs, we have shown that the astrometric performances are still exceptional in the
central arcminute of the simulated GCs: most of the stars have errors around a
few 100~$\mu$as or $\mu$as~yr$^{-1}$.}
\item{The BP/RP photometry and the RVS spectra, on the other hand, have larger AL
sizes and therefore suffer more from crowding effects. The most important factor
for these intruments is crowding by field stars, especially in the most extreme
cases like the bulge field, which acts at all distances from the GC center and
combines with the crowding effects from GC stars in the central regions.}
\end{enumerate}

The imminent decision of whether to extend the Gaia mission lifetime will
certainly have a beneficial impact on all the above measurements. However, the
simulations presented here are already pessimistc and the pipelines are expected
to evolve significantly in the next few years. Therefore, we conclude that Gaia
measurements will revolutionize our kowledge of GCs.

\section*{Acknowledgements}

We warmly thank A.~Brown, M.~Castellani, A.~Di Cecco, F.~De Luise, D.~Harrison,
H.~E.~Huckle, K.~Janssen, C.~Jordi, P.~M.~Marrese, D.~Pourbaix, L.~Pulone, and
G.~Seabroke, for providing useful information about Gaia BP, RP, and RVS
deblending and decontamination, and for enlightening discussions about Gaia and
crowded fields treatment in general. This work used simulated data provided by
the Simulation Unit (CU2) of the Gaia Data Processing Analysis Consortium
(DPAC), run with GIBIS at CNES (Centre national d'\'etudes spatiales). This
research made use of the R programming language (https://www.r-project.org/) and
more specifically of its `data.table' package, for the treatment of very large
datasets. Some of the figures were prepared with TopCat \citep{topcat}.

\appendix

\section{List of acronyms}
\label{sec:acr}

Table~\ref{tab:acr} lists all the acronyms used.

\begin{table}
\caption{List of acronyms used in this paper. \label{tab:acr}}
\begin{center}
\begin{tabular}{ll}
\hline
\\
Acronym   & Description \\
\hline
AC        & ACross scan \\
AF        & Astrometric field \\
AL        & ALong scan \\
BP        & Blue spectro-Photometer \\
CCD       & Charge Coupled Device \\
CDS       & Centre de Donn\'ees astronomiques de Strasbourg \\
CMD       & Color Magnitude Diagram \\
CNES      & Centre National d'\'Etudes Spatiales \\
DPAC      & Data Processing and Analysis Consortium \\
ESA       & European Space Agency \\
FWHM      & Full Width at Half Maximum \\
GIBIS     & Gaia Instrument and Basic Angle Simulator \\
GC        & Globular Cluster \\
HB        & Horizontal Branch \\
HST       & Hubble Space Telescope \\
IMF       & Initial Mass Function \\
LSF       & Line Spread Function \\
MLE       & Maximum Likelihood Estimator \\
MW        & Milky Way \\
NSS       & Non Single Stars \\
PSF       & Point Spread Function \\
RAVE      & RAdial Velocity Experiment \\
RP        & Red spectro-Photometer \\
RV        & Radial Velocity \\
RVS       & Radial Velocity Spectrometer \\
SEA       & source Environment Analysis \\
SM        & Sky Mapper \\
TDI       & Time Delayed Integration \\
\hline
\end{tabular}\\
\end{center}
\end{table}

\label{lastpage}
\end{document}